\documentclass[reprint, amsmath, amssymb, aps, prl, superscriptaddress, nobibnotes, longbibliography]{revtex4-2}

\usepackage{soul}
\usepackage{siunitx}

\usepackage{graphicx}
\usepackage{braket}
\usepackage[normalem]{ulem}
\usepackage[colorlinks]{hyperref}

\begin{document}

\title{Efficient Source of Shaped Single Photons Based on an Integrated Diamond Nanophotonic System}

\author{E. N. Knall}
\thanks{These authors contributed equally to this work.}
\affiliation{John A. Paulson School of Engineering and Applied Sciences, Harvard University, Cambridge, Massachusetts 02138, USA}
\author{C. M. Knaut}
\thanks{These authors contributed equally to this work.}
\affiliation{Department of Physics, Harvard University, Cambridge, Massachusetts 02138, USA}
\author{R. Bekenstein}
\thanks{These authors contributed equally to this work.}
\affiliation{Department of Physics, Harvard University, Cambridge, Massachusetts 02138, USA}
\affiliation{Racah Institute of Physics, The Hebrew University of Jerusalem, Jerusalem 91904, Israel}
\author{D. R. Assumpcao}
\thanks{These authors contributed equally to this work.}
\affiliation{John A. Paulson School of Engineering and Applied Sciences, Harvard University, Cambridge, Massachusetts 02138, USA}
\author{P. L. Stroganov}
\affiliation{Department of Physics, Harvard University, Cambridge, Massachusetts 02138, USA}
\author{W. Gong}
\affiliation{Department of Physics, Harvard University, Cambridge, Massachusetts 02138, USA}
\author{Y. Q. Huan}
\affiliation{Department of Physics, Harvard University, Cambridge, Massachusetts 02138, USA}
\author{P. -J. Stas}
\affiliation{Department of Physics, Harvard University, Cambridge, Massachusetts 02138, USA}
\author{B. Machielse}
\affiliation{Department of Physics, Harvard University, Cambridge, Massachusetts 02138, USA}
\affiliation{AWS Center for Quantum Computing, Pasadena, California 91125, USA}
\author{M. Chalupnik}
\affiliation{Department of Physics, Harvard University, Cambridge, Massachusetts 02138, USA}
\author{D. Levonian}
\affiliation{Department of Physics, Harvard University, Cambridge, Massachusetts 02138, USA}
\affiliation{AWS Center for Quantum Computing, Pasadena, California 91125, USA}
\author{A. Suleymanzade}
\affiliation{Department of Physics, Harvard University, Cambridge, Massachusetts 02138, USA}
\author{R. Riedinger}
\affiliation{Department of Physics, Harvard University, Cambridge, Massachusetts 02138, USA}
\affiliation{Institut f\"ur Laserphysik und Zentrum f\"ur Optische Quantentechnologien, Universit\"at Hamburg, 22761 Hamburg, Germany}
\affiliation{The Hamburg Centre for Ultrafast Imaging, 22761 Hamburg, Germany}
\author{H. Park}
\affiliation{Department of Physics, Harvard University, Cambridge, Massachusetts 02138, USA}
\affiliation{Department of Chemistry and Chemical Biology, Harvard University, Cambridge, Massachusetts 02138, USA}
\author{M. Lon\v{c}ar}
\affiliation{John A. Paulson School of Engineering and Applied Sciences, Harvard University, Cambridge, Massachusetts 02138, USA}
\author{M. K. Bhaskar}
\affiliation{Department of Physics, Harvard University, Cambridge, Massachusetts 02138, USA}
\affiliation{AWS Center for Quantum Computing, Pasadena, California 91125, USA}
\author{M. D. Lukin}
\affiliation{Department of Physics, Harvard University, Cambridge, Massachusetts 02138, USA}
\date{June 2022}

\begin{abstract}
    An efficient, scalable source of shaped single photons that can be directly integrated with optical fiber networks and quantum memories is at the heart of many protocols in quantum information science. We demonstrate a deterministic source of arbitrarily temporally shaped single-photon pulses with high efficiency (detection efficiency = 14.9\%) and purity ($g^{(2)}(0) = 0.0168$) and streams of up to 11 consecutively detected single photons using a silicon-vacancy center in a highly directional fiber-integrated diamond nanophotonic cavity. Combined with previously demonstrated spin-photon entangling gates, this system enables on-demand generation of streams of correlated photons such as cluster states and could be used as a resource for robust transmission and processing of quantum information.
\end{abstract}

\maketitle

Single optical photons play an essential role in quantum information tasks ranging from quantum communication \cite{Kimble2008} to measurement-based quantum computing \cite{Raussendorf2001,Kok2007}. 
Many protocols in quantum communication use single photons as information carriers between remote locations since photons experience little decoherence while propagating in an optical fiber or free space over long distances. An efficient, scalable source of single photons is therefore extremely useful in quantum information science and technology \cite{Wang2019, Eisaman2011}.

The most promising approaches for realizing single-photon sources are based on single atoms, ions \cite{Kuhn2002,McKeever2004,Keller2004}, or artificial atoms \cite{Michler2000,Yuan2002, Beveratos2002} coupled to optical cavities. The underlying idea is that by promoting an atom to its excited state in a controlled way, only one photon is emitted per excitation cycle. Meanwhile, the encapsulating optical cavity ensures a high probability of photon collection into a well-defined optical mode. Numerous state-of-the-art demonstrations of single-photon sources have utilized solid-state, cavity-integrated self-assembled quantum dots \cite{Uppu2020,He2017,Unsleber2016,Wang2019,Arcari2014}, which have recently been used in an experiment demonstrating in-fiber single-photon detection efficiencies of above 50\% \cite{Tomm2021}. 

However, in addition to single photons and linear optical elements, key quantum communication applications such as complex quantum networks will eventually require the use of more advanced components such as quantum memories and quantum repeaters to correct loss errors in communication channels \cite{Briegel1998, Muralidharan2014} or serve as a deterministic nonlinearity to enable quantum logic gates between itinerant photons \cite{Kalb2015,Duan2004}. The necessity of integrating single photons with other components of future quantum networks creates additional requirements that many present-day single-photon sources do not meet: control over the photon frequency, bandwidth, and temporal profile. In particular, leading quantum memory systems have limited bandwidths, often on the MHz scale, which is several orders of magnitude smaller than the bandwidths of typical state-of-the-art single-photon sources \cite{Baek2008,Ma2011}. While bandwidth-tailored sources have been realized with neutral-atoms \cite{Mucke2013}, trapped ions \cite{Schupp2021}, and quantum dots \cite{Pursley2018, He2013}, such systems with high end-to-end efficiencies, compatibility with scalable device fabrication, and photonic integration have yet to be demonstrated.

In this Letter, we present a versatile, fiber-coupled single-photon source based on a silicon-vacancy center in diamond which features high efficiency, purity, temporal control, integrability, and access to auxiliary spin memory registers. It can also directly interface with existing quantum memories, enabling future compatibility with repeater-based quantum networks as well as protocols for generation of streams of entangled photonic graph states, which are key resources in rapid one-way quantum communication and measurement-based quantum computation protocols \cite{Borregaard2020,Schwartz2016,Economou20102, Tiurev2020, Buterakos2017, Lindner2009}.

\begin{figure}
    \centering
    \includegraphics[width=\linewidth]{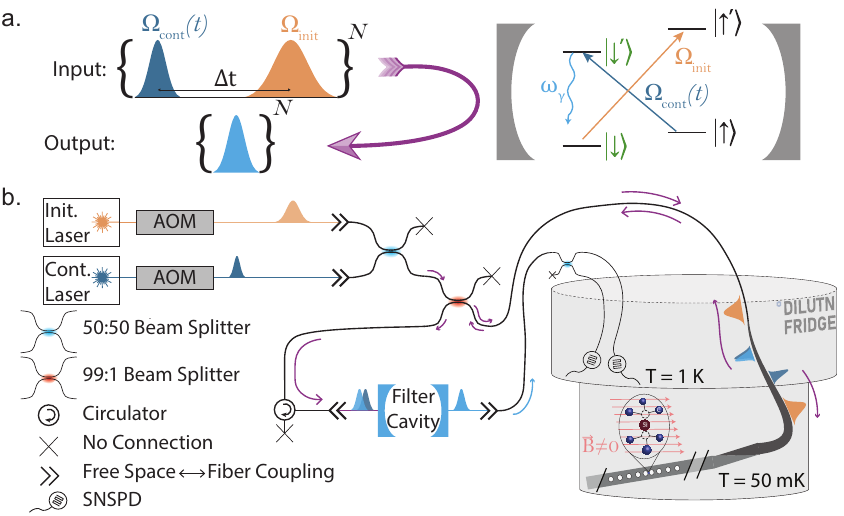}
    \caption{(a) Photon Creation Schematic. The four-level system of the SiV spin is coherently driven by alternating initialization ($\Omega_\text{init}$) and photon generation ($\Omega_\text{cont}$) optical pulses, producing a train of temporally shaped photons which are efficiently collected with an overcoupled nanophotonic cavity. (b) Measurement Setup Schematic. A nanophotonic cavity containing an SiV is cooled to 50 mK in a dilution refrigerator and pumped to coherently create single, arbitrarily-shaped photons. Pump pulses are shaped using an acousto-optic modulator, and pump light is filtered out of the single-photon stream by a free space Fabry-–P\'erot cavity.}
    \label{fig:1}
\end{figure}

Our system consists of a single negatively charged silicon-vacancy center (SiV) in a diamond nanophotonic cavity. The SiV is an inversion-symmetric point defect which features an optically accessible quantum memory that can be embedded in nanofabricated structures while maintaining excellent spin and optical coherence \cite{Nguyen2019, Nguyen2019a}. Our cavity-quantum electrodynamics (CQED) system exhibits strong light-matter coupling, characterized by the single-photon Rabi frequency and cavity and atomic energy decay rates $\{g, \kappa, \gamma\} = 2 \pi \times \{\SI{6.81}{\giga\hertz}, \SI{329}{\giga\hertz}, \SI{0.1}{\giga\hertz} \}$, resulting in a cooperativity of $C \approx 6$ \cite{SI}(I.4.2). Unlike in previous experiments where the magnetic field was oriented along the main symmetry axis of the SiV, we apply a magnetic field nearly orthogonal to the SiV's symmetry axis, giving rise to a four-level system corresponding to the ground ($\ket{\downarrow}, \ket{\uparrow}$) and optically excited ($\ket{\downarrow '}, \ket{\uparrow '}$) states of the SiV's electronic hole spin \cite{Hepp2014,Hepp2014a}. The orthogonal field orientation results in spin-flipping optical transitions becoming allowed, hence enabling fast spin initialization and photon generation \cite{Sukachev2017, Rogers2014}.

The protocol for single-photon generation in this system is illustrated schematically in Fig. \ref{fig:1}a. First, the four-level system is initialized in $\ket{\uparrow}$ by optically pumping the spin flipping transition $\ket{\downarrow} \rightarrow \ket{\uparrow'}$ using a classical driving field with Rabi frequency $\Omega_{\text{init}}$. Then, the population is coherently transferred to a single photon using a control pulse with Rabi frequency $\Omega_{\text{cont}}$ to drive the transition $\ket{\uparrow} \rightarrow \ket{\downarrow'}$. Repeated application of this pulse sequence generates streams of single photons.

The temporal profile of the single-photon wavepackets can be tuned on timescales much longer than the excited state $\ket{\downarrow'}$ lifetime due to the long-lived quantum memory of the SiV spin. In the limit of weak driving $|\Omega_\text{cont}| \ll \Gamma$, where $\Gamma$ is the cavity-enhanced decay rate along $\ket{\downarrow '} \rightarrow \ket{\downarrow}$, the dynamics of the excited state $\ket{\downarrow^{\prime}}$ adiabatically follows the excitation process \cite{SI}(II) and the photon linewidth is limited only by the coherence of the spin-levels $\{\ket{\downarrow}, \ket{\uparrow}\}$ and the control laser's linewidth, rather than the intrinsic lifetime of the SiV excited state $\ket{\downarrow '}$. By modulating the strength and shape of the control pulse $\Omega_\text{cont}(t)$, we temporally shape the single photon.

The schematic of our experimental setup is shown in Fig.~\ref{fig:1}b. Devices are placed in a dilution refrigerator at $T \approx 50$ mK to reduce the population of phonons which cause thermal mixing between orbital states \cite{Sukachev2017, Jahnke2015}. This extends the coherence of the ground-state spin, enabling generation of temporally longer photon pulse shapes. Optical control pulses are delivered, and single photons are collected via a tapered optical fiber, which is coupled to the device \cite{Burek2017}. On the return path, the generated single photons are filtered from the control pulses by a free space Fabry-–P\'erot cavity (linewidth = 160 MHz, finesse = 312) before being detected by superconducting nanowire single-photon detectors (SNSPDs).

\begin{figure}
    \centering
    \includegraphics[width=\linewidth]{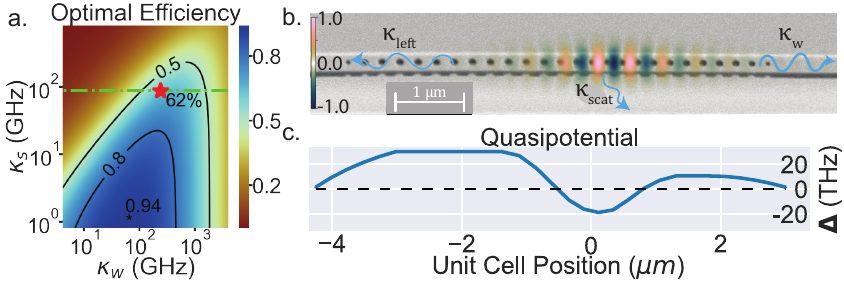}
    \caption{(a) Photon extraction efficiency is shown in the color plot as a function of cavity-waveguide coupling $\kappa_{\text{w}}$ and unwanted cavity loss rate $\kappa_{\text{s}}$. Contours aid readability of color map. Optimal extraction efficiency is maximized by trading off atom-photon interaction probability, proportional to $(\kappa_{\text{s}} + \kappa_{\text{w}})^{-1}$, for a higher cavity-waveguide coupling rate $\kappa_{\text{w}}$. The dashed-dot line cut corresponds to $\kappa_{\text{s}} = 89$ GHz, the unwanted loss rate of this device, which is determined by fabrication imperfections. The red star highlights this device with waveguide coupling rate $\kappa_{\text{w}} = 240$ GHz, which is nearly optimal for the given $\kappa_{\text{s}}$. (b) A scanning electron micrograph (SEM) of the nanophotonic cavity is overlaid with the simulated electric field, and loss rates are labeled, where $\kappa_{\text{s}} = \kappa_\text{scat} + \kappa_\text{left}$. (c) The simulated quasipotential shape of the cavity \cite{SI}(I.1) shows that there is a lower and shorter potential barrier on the weak mirror side. This corresponds to the coupling to the right waveguide being the dominant loss rate ($\kappa_{\text{w}} \gg \kappa_{\text{s}}$).}
    \label{fig:2}
\end{figure}
 
 \begin{figure*}
    \centering
    \includegraphics[width=\textwidth]{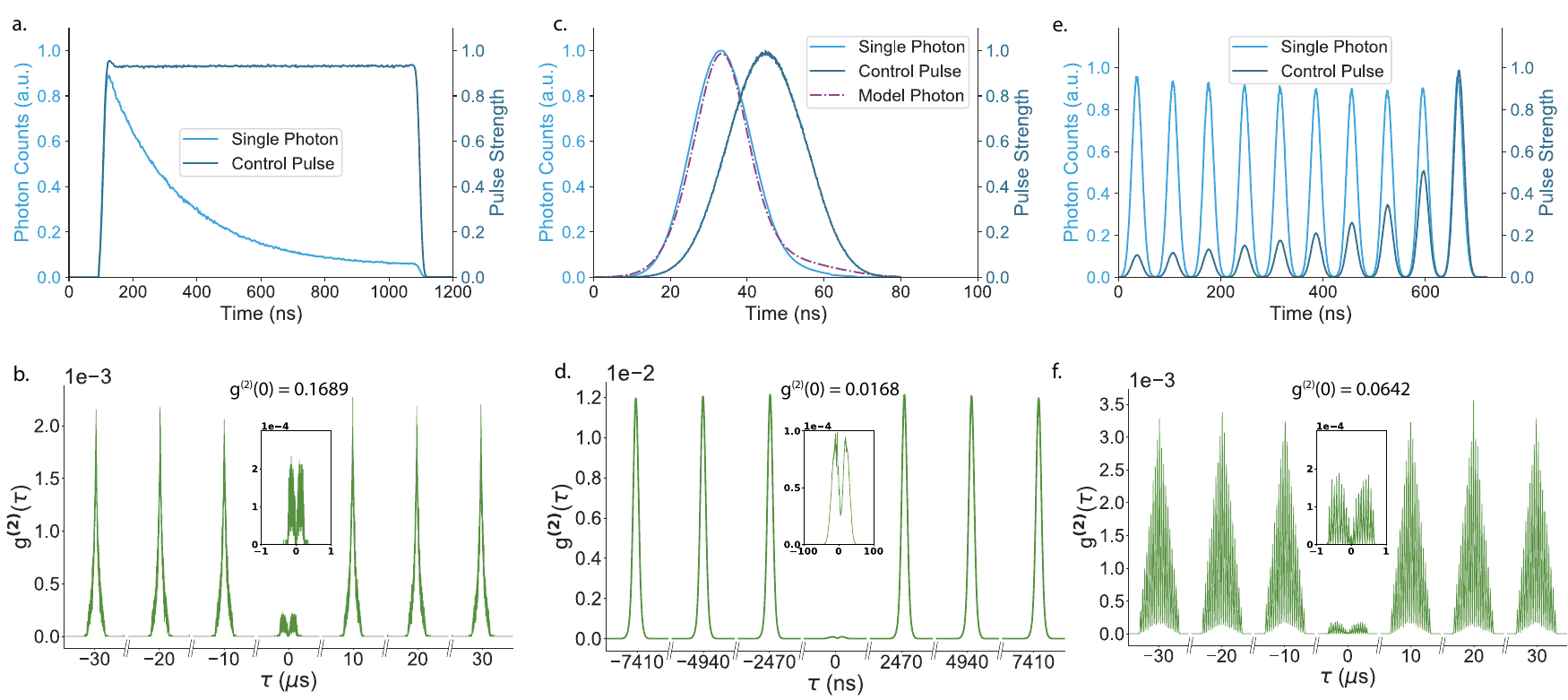}
    \caption{Pulse-shaped single-photon generation. Upper panels (a), (c), (e) display the temporal profile of the coherent control pulse and detected single photon. (a) A square control pulse produces an exponentially decaying photon. (b) A Gaussian single photon. (c) A single photon distributed over ten time bins (e). Lower panels (b), (d), (f) display the normalized second order correlation of photon arrivals for the exponential, Gaussian, and ten peaked photons respectively. The insets show a zoomed-in window around $\tau = $ 0, which is integrated to calculate $g^{(2)}(0)$.}
    \label{fig:3}
\end{figure*}

In order to maximize the photon collection efficiency from the emitter, we implement a novel asymmetric nanophotonic cavity design which strikes a balance between high quality factor of the cavity and strong waveguide damping. In this design, for a given unwanted cavity loss rate $\kappa_\text{s}$ set by fabrication imperfections, there is an optimal choice of waveguide coupling $\kappa_\text{w}$ (Fig. \ref{fig:2}a). We achieve this optimal trade-off in the asymmetric diamond nanophotonic cavity as pictured in the scanning electron micrograph (SEM) of Fig. \ref{fig:2}b, which preferentially sends light to the coupling waveguide (i.e. to the right side).

These devices are designed using the analogy between a massive particle tunneling through a potential barrier and the evanescent decay of a photon in a photonic band gap \cite{Joannopoulos2008}. The asymmetric ``quasipotential" for a photon in this device is shown in Fig. \ref{fig:2}c. It illustrates both the preferential coupling to the measurement port through the lower and narrower barrier on the right side of the cavity as well as a deep well needed for the tight confinement of the optical mode. The simulated electric field overlay in Fig. \ref{fig:2}b illustrates this wavelength scale confinement (mode volume = 0.67 $\left(\frac{\lambda}{n}\right)^3$). An in-depth discussion of the new photonic crystal cavity design techniques used here is provided in section I of the Supplementary Information \cite{SI}.

\begin{figure}
    \centering
    \includegraphics[width=\linewidth]{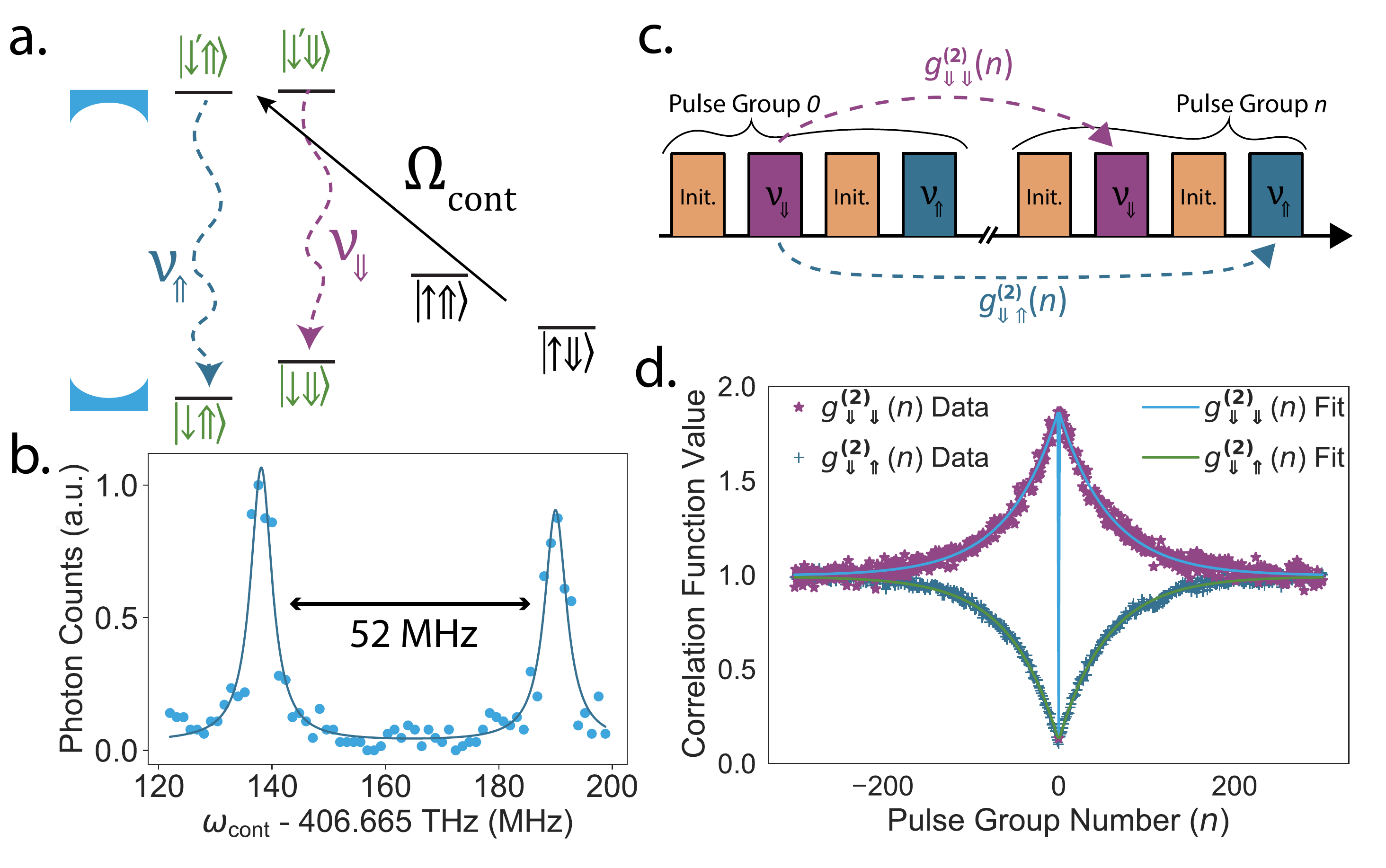}
    \caption{Hyperfine splitting due to the $^{29}\mathrm{Si}$ nuclear spin (a) gives rise to a four-level ground state manifold. Pumping on the electron-spin flipping transition with $\Omega_{\text{cont}}$ results in the generation of photons with two nuclear-state dependent frequencies $\nu_{\Uparrow}$ and $\nu_{\Downarrow}$. (b) Sweeping the control pulse frequency selectively tunes $\nu_{\Uparrow}$ and $\nu_{\Downarrow}$ into resonance with the filter cavity, which enables the measurement of the spectrum of the emitted photons. (c) Pulse sequence for attempting to measure either two consecutive photons at nuclear-state dependent frequencies $\nu_{\Downarrow}$ and $\nu_{\Downarrow}$ or at nuclear-state dependent frequencies $\nu_{\Downarrow}$ and $\nu_{\Uparrow}$. (d) (Purple) ${g_{\Downarrow\Downarrow}}^{(2)}(\tau)$ auto-correlation measurements of the photons emitted at $\nu_{\Downarrow}$ show antibunching at zero time delay, and bunching after emission of $113.8 \pm 3.8$ photons ($1.48$ ms timescale). (Blue) ${g_{\Downarrow \Uparrow}}^{(2)}(\tau)$ cross-correlation function for the two consecutively emitted photons at $\nu_{\Uparrow}$ and $\nu_{\Downarrow}$ shows anti-bunching after emission of $110.6 \pm 2.4$ photons, suggesting that the nuclear polarization is preserved for repeated generation of up to 110 photons.
    }
    \label{fig:5}
\end{figure}

We demonstrate generation of bandwidth-tailored photons with this platform in Fig. \ref{fig:3}. We start by applying a $1~\mu$s square pump pulse (Fig.~\ref{fig:3}a), and observing an exponentially shaped emitted photon, directly illustrating the optical pumping dynamics from $\ket{\uparrow}$ into $\ket{\downarrow}$ expected from a time-independent pump pulse. The photon duration of $\sim 1~\mu$s compared to the $\sim 1$ ns excited state lifetime highlights the ability of this protocol to generate narrow-bandwidth photons, which is independently verified using a separate narrow filter cavity with linewidth below 5 MHz (Fig \ref{fig:5}b). To ensure the presence of only a single photon in each wavepacket, we measure the second order correlation, $g^{(2)}$, of the generated photon stream by using a beam splitter and recording the arrival times of photons on a pair of SNSPDs. The results of these measurements are shown in the lower panels of Fig. \ref{fig:3}. A value of $g^{(2)}(0) = 0.1689 < 0.5$ (Fig.~\ref{fig:3}b) of the exponentially shaped photon confirms the quantum nature of the measured state of light and the presence of a single excitation.

Next, we apply shorter and more powerful Gaussian control pulses to create Gaussian single photons with full-width half-maxima of $\sim$ $20$ ns (Fig. \ref{fig:3}c) and observe a substantially reduced $g^{(2)}(0) = 0.0168$ (Fig. \ref{fig:3}d). We note that photons of this approximate duration are optimal for interfacing with existing SiV quantum memories \cite{Bhaskar2020, SI}.

We confirm our understanding of the system using a density matrix model \cite{SI}(III) to predict the photon shapes resulting from the applied Gaussian control pulse. Fig. \ref{fig:3}c confirms that our model matches the measured photon shape well. By inverting this model, we can calculate the control pulse $\Omega_{\text{cont}}(t)$ required to generate arbitrarily shaped photon wavepackets. For example, in Fig. \ref{fig:3}e, we demonstrate a ten-peaked single photon, which could be useful for time-binned multiplexing \cite{Bhaskar2020} and efficient high-dimensional quantum communication \cite{MiguelRamiro2018,Cozzolino2019}. Auto-correlation measurements again demonstrate the single-photon nature of the ten-peaked photon with a low $g^{(2)}(0) = 0.0642$. The difference in single-photon purity between the three generated photon shapes can be attributed to optically-induced heating, which results in a reduced spin lifetime and increased value of $g^{(2)}(0)$ for longer-duration photons \cite{SI}(VI.2).

\begin{figure}
    \centering
    \includegraphics[width=\linewidth]{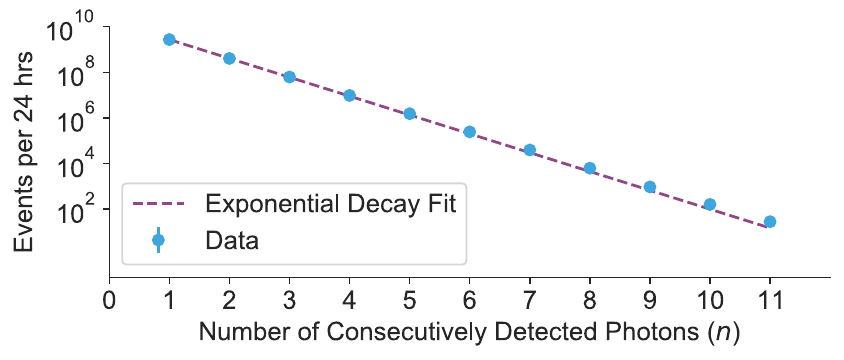}
    \caption{Statistics of consecutive $n$-photon streams detected during a 24-hour acquisition at a 405 kHz repetition rate and 57\% average duty cycle, showing detection of up to 11 photons in a row. Exponential decay fit indicates a total source-to-detector efficiency of $14.9\%$.}
    \label{fig:4}
\end{figure}

Next, we measure the total system efficiency by generating short Gaussian photons (as in Fig. \ref{fig:3}c) continuously over a 24 hour period. The repetition rate of the pump pulses is 405 kHz. We record the number of consecutive $n$-photon streams detected (Fig. \ref{fig:4}) as a proxy for the complexity of multi-photon states that are necessary for implementation of quantum information protocols such as one-way quantum communication or computing with photonic cluster states \cite{Borregaard2020,Schwartz2016,Economou20102, Tiurev2020, Buterakos2017, Lindner2009}. Notably, the experiment was operating 
autonomously during this 24-hour run. Our experiment control software \cite{pylabnet} automatically handles SiV ionization and spectral diffusion events, as well as filter cavity locking \cite{SI}(IV.2), making this a realistic demonstration of a practical single-photon source.

The exponential decay fit to the $n$-photon event rates reveals a single-photon detection efficiency of $14.9\%$ \cite{SI}(V.1). This decrease compared to the ideal photon extraction efficiency of $62 \%$ is primarily due to losses in the filtering setup (0.5-0.6), waveguide-fiber coupling efficiency (0.7), and finite detuning of the cavity \cite{SI}(IV.3). Despite these extra losses, this single-photon efficiency is competitive with state-of-the-art single-photon sources \cite{Uppu2020,Kirsanske2017,Wang2019}. A single-photon detection rate of 31 kHz is achieved, indicating an average duty cycle of 57\%, which is primarily limited by ionization of the SiV and software overhead \cite{SI}(V.2). 

As a first step toward generation of spin-photon entangled states and more complex multi-photon entangled states, we next explore the light-matter interface with the auxiliary nuclear spin memory associated with the $^{29}$Si isotope. The hyperfine coupling between the electronic hole spin and the nuclear spin additionally splits the electronic-hole Zeeman ground state manifold creating four levels in the ground-state manifold (Fig. \ref{fig:5}a). As a result, the $^{29}$SiV system can emit photons with two nuclear spin dependent photon frequencies, $\nu_{\Uparrow}$ and $\nu_{\Downarrow}$. Such a system can be used to generate complex multi-photon entangled states such as cluster states or graph states, as proposed in \cite{Tiurev2020, Lindner2009}, by coherently manipulating the nuclear state in between emissions of subsequent photons.

In order to probe the nuclear spin dependent emission frequency of a cavity-integrated $^{29}$SiV, we filter the single-photon signal using a significantly narrower 5 MHz linewidth filter cavity, locked close to the $ \ket{\downarrow '} \rightarrow \ket{\downarrow}$ transition. The photons are generated via the same scheme as before, whereby a single initialization pulse is used to initialize the electron regardless of the initial nuclear spin state due to the small hyperfine splitting as compared to the optical transition bandwidth. The filter cavity frequency is held constant while the frequency of the pump laser is swept tuning the frequency of the emitted photons. This selectively tunes $\nu_{\Uparrow}$ and $\nu_{\Downarrow}$ into resonance with the filter cavity, enabling the measurement of the spectrum of the emitted photons. Two narrow peaks are observed in the detected single-photon spectrum with a splitting of 52 MHz (Fig.~\ref{fig:5}b), as expected from the hyperfine splitting from the $^{29}\mathrm{Si}$ nuclear spin \cite{Pingault2017}.

An initial step toward generating multi-photon states with entanglement mediated by the $^{29}$SiV nuclear spin is to show that multiple photons can be generated while preserving the nuclear spin state. Therefore, we measure correlations between subsequently emitted photons at the two different nuclear spin dependent emission frequencies, $\nu_{\Uparrow}$ and $\nu_{\Downarrow}$ (Fig. \ref{fig:5}c). We measure the degree of second-order correlations $g_{\Downarrow \Downarrow}^{(2)}(\tau)$ of photons emitted at frequency $\nu_{\Downarrow}$, observing bunching on long timescales. We then measure the intensity cross-correlation ${g_{\Downarrow \Uparrow}}^{(2)}(\tau) = \langle I_{\Downarrow}(t) I_{\Uparrow}(t+\tau) \rangle / \langle I_{\Downarrow}(t) I_{\Uparrow}(t) \rangle$, where $ I_{\Uparrow}$ and $ I_{\Downarrow}$ are the intensities of the $\nu_{\Uparrow}$ and $\nu_{\Downarrow}$ emissions, respectively, and observe anti-bunching on the same timescale. These measurements indicate a 16-fold higher probability of detecting subsequently emitted photons at the same frequency, as opposed to opposite frequencies.

The bunching (antibunching) in $g_{\Downarrow\Downarrow}^{(2)}(\tau)$ (${g_{\Downarrow\Uparrow }}^{(2)}(\tau)$) decays after emission of $113.8 \pm 3.8$ ($110.6 \pm 2.4$) photons. We attribute this decay to relaxation of the nucleus due to the single-photon generation process. Relaxation of the nucleus after emission of $113.8 \pm 3.8$ photons would correspond to each generated photon inducing a nuclear spin flip with a probability of $(0.9 \pm 0.03) \%$. These measurements directly demonstrate that classical correlations between the $^{29}\mathrm{Si}$ nuclear spin state and the frequency of the emitted photon can persist for more than 100 consecutively emitted photons, making this a promising approach for the generation of large-scale photonic graph states \cite{Lindner2009}.

Our experiments demonstrate an on-demand source of streams of shaped photons generated from a silicon-vacancy center in an asymmetric nanophotonic cavity in diamond. The challenge of producing a nanophotonic cavity in diamond with arbitrary coupling ratios was resolved through the development of a quasipotential design heuristic, which we believe will be of general use to the nanophotonics community. We showed that the system can generate single photons with highly tunable temporal wavepackets and high spectral purity, detecting streams of up to 11 sequential photons at experimentally useful rates due to a high source-to-detector efficiency and efficient fiber-nanophotonic integration. Given the measured $g^{(2)}(0) = 0.0168$, we estimate this source would provide more than a thirty-fold improvement in the single-photon detection rate when used as a replacement for a weak coherent source with equivalent two-photon detection infidelity \cite{SI}(V.3). Furthermore, this advantage results in an exponential improvement for higher $n$-photon stream events, as demonstrated by the detection of 28 total 11-photon events in a 24 hour period, which is comparable to state-of-the-art \cite{Zhong2018}.

Additionally, this single-photon source should enable the generation of multi-photon entangled states when efficiently interfaced with a second cavity-coupled SiV \cite{Bhaskar2020}, which would be used as a quantum memory to deterministically entangle subsequent photons \cite{Lindner2009}. By demonstrating classical correlations between the built-in $^{29}$Si nuclear spin state and emitted photon frequency, we also illustrate the possibility to directly generate streams of entangled photons mediated by nuclear memory. In order to demonstrate quantum correlations (i.e. entanglement) between nuclear spin and photon frequency, additional coherent control of the nucleus would be necessary, which should be possible using RF fields supplied by on-chip coplanar waveguides \cite{Nguyen2019, Nguyen2019a}. Moreover, in order to realize large entangled states, mitigation of $^{29}\mathrm{Si}$ memory decoherence arising from heating, which shortens the electron lifetime, will be required \cite{SI}(VI.1).

This work builds on our previous demonstration of the SiV-based quantum memory node \cite{Bhaskar2020}. The photons generated by our source can be bandwidth and wavelength matched to existing SiV-nanophotonic quantum memory devices, which will be required for complex quantum networking schemes involving stationary repeaters or quantum memories. Combined with the demonstrated ability to create large multi-photon streams on demand, this method should enable the production and detection of high photon number linear cluster states with only moderate improvements to the setups demonstrated here and in \cite{Bhaskar2020, SI}. For these reasons, our platform demonstrates promise as a versatile single-photon source which can be interfaced with quantum memories for the realization of quantum networking and quantum information processing tasks.

The authors thank Denis Sukachev and Yan-Cheng Wei for their insightful discussions and feedback on the manuscript, as well as Jim MacArthur
for assistance with electronics. This work was supported by the NSF (PHY-2012023), NSF EFRI ACQUIRE (5710004174), CUA (PHY-1734011), DoE (DE-SC0020115), AFOSR MURI (FA9550171002 and FA95501610323), and CQN (EEC-1941583). Devices were fabricated in the Harvard University Center for Nanoscale Systems (CNS), a member of the National Nanotechnology Coordinated Infrastructure Network (NNCI), which is supported by the National Science Foundation under NSF award no. 1541959. ENK, DA, and BM acknowledge that this material is based upon work supported by the National Science Foundation Graduate Research Fellowship under Grant No. DGE1745303. YQH acknowledges support from the Agency for Science, Technology and Research (A*STAR) Singapore through the National Science Scholarship. MC and MKB acknowledge support from the Department of Defense (DoD) through the National Defense Science and Engineering Graduate (NDSEG) Fellowship Program. RR acknowledges support from the Alexander von Humboldt Foundation and the Cluster of Excellence `Advanced Imaging of Matter' of the Deutsche Forschungsgemeinschaft (DFG) - EXC 2056 - project ID 390715994. The color maps used in Fig. \ref{fig:2} were designed by \cite{Crameri2020}.

\nocite{Lodahl2015,Lobet2020,Quan2010a,Evans2018,Kastoryano2011,Borregaard2015,RevModPhys.87.1379,Tiecke2014,weisskopf1997berechnung,qutip2,Senellart2017,srujanstrain,Machielse2019,Akahane2003,pylabnet}

\bibliographystyle{apsrev4-2}
\bibliography{refs}

\end{document}


\title{Supplementary Information for ``Efficient Source of Shaped Single Photons Based on an Integrated Diamond Nanophotonic System"}

\date{May 2022}

\maketitle

\section{Nanophotonic Cavity}
\label{section:nanophotonics}
In order to be widely useful, the high-purity single photons must be generated and outcoupled efficiently, ideally into a single-mode fiber. This is especially true when generating large multi-photon states for which generation probability scales exponentially with single-photon efficiency. Single photon generation and outcoupling efficiency for an atom-cavity system $\eta$ is determined by both the atom-cavity cooperativity $C$ as in $p_{c} = \frac{C}{C+1}$ and the overcoupling of the cavity to the waveguide as in $p_{w} = \frac{\kappa_{w}}{\kappa_w + \kappa_s}$, as well as additional system losses from imperfect optical elements:
\begin{equation} \label{eff}
    \eta = p_{c}*p_{w}*p_{o} 
\end{equation}
Where $p_{c}$ is the probability of emission into the cavity, and $p_{w}$ is the probability that a photon in the cavity couples into the probe waveguide. Systemic optical losses ($p_o$), although crucial in practical systems and often dependent on the physical platform, are not fundamental to the cavity quantum electrodynamics (CQED) system design, so we do not include them in our optimization of source efficiency $\eta_{s}$. Instead, we focus on the first two terms, $\eta_{s}= p_{c}*p_{w}$.

When the emitter is close to resonance with the cavity, the probability of emission of a single photon into the cavity is given by the cooperativity as $p_{c} = \frac{C}{C+1}$ \cite{RevModPhys.87.1379} with $C = \frac{4g^2}{\kappa\gamma}$ where $g$ is the atom-cavity coupling rate, $\kappa$ is the full width at half max (FWHM) cavity energy decay rate, and $\gamma$ is the quantum emitter's spontaneous decay rate (FWHM). Meanwhile, the probability that a photon in the cavity will decay into the waveguide is given by $p_{w} = \frac{\kappa_{w}}{\kappa_w + \kappa_s}$, where $\kappa_s$ represents all unwanted cavity loss, such as scattering into free-space or coupling into the unprobed waveguide. In terms of CQED parameters, the source efficiency is
\begin{equation} \label{CQED_efficiency}
    \eta_{s}= \frac{4 g^2 \kappa_w}{(\kappa_{s}+\kappa_w)(4g^2+(\kappa_{s}+\kappa_w)\gamma)}
\end{equation}
Maximizing this function optimizes the photon generation efficiency. Thus, the optimal decay rate into the waveguide is given by
\begin{equation} \label{k_opt}
    \kappa_w^{(opt)} = \sqrt{\frac{\kappa_{s}(4g^2+\kappa_{s}\gamma)}{\gamma}}
\end{equation}

Notably, device performance, in this case single-photon generation efficiency, is not maximized by simply maximizing device cooperativity, as is often the case in other CQED protocols \cite{Kastoryano2011,Borregaard2015,Evans2018,Bhaskar2020}. In fact, as illustrated by equation \ref{k_opt} and fig. 2a in the main text, optimal overall photon generation efficiency is achieved by trading off some cooperativity for a better balance between relative cavity coupling rates $\kappa_{w}$ and $\kappa_{s}$. This is distinct from earlier diamond nanophotonic cavities which were designed to maximize cooperativity and be nearly critically coupled in order to maximize spin-dependent reflection contrast.\cite{Bhaskar2020} This underscores the importance of creating a cavity with well-controlled coupling rates.

The nanophotonic cavities used in this experiment are similar to those used in the previous CQED experiments with silicon-vacancy centers in diamond (SiV) \cite{Bhaskar2020,Nguyen2019,Nguyen2019a,Evans2018}. However, while previous generations of cavities were designed to maximize cooperativity (or equivalently Purcell enhancement), these nanophotonic cavities were designed to balance maximizing cooperativity with ensuring directional waveguide coupling. This additional constraint forced us to expand the design parameter space. The resulting design is asymmetric throughout the cavity, which is different from traditional photonic crystal cavity design. We discuss the subtle but fundamental difference below after some brief background on photonic crystal cavities.

\begin{figure}
    \centering
    \includegraphics[width=\linewidth/2]{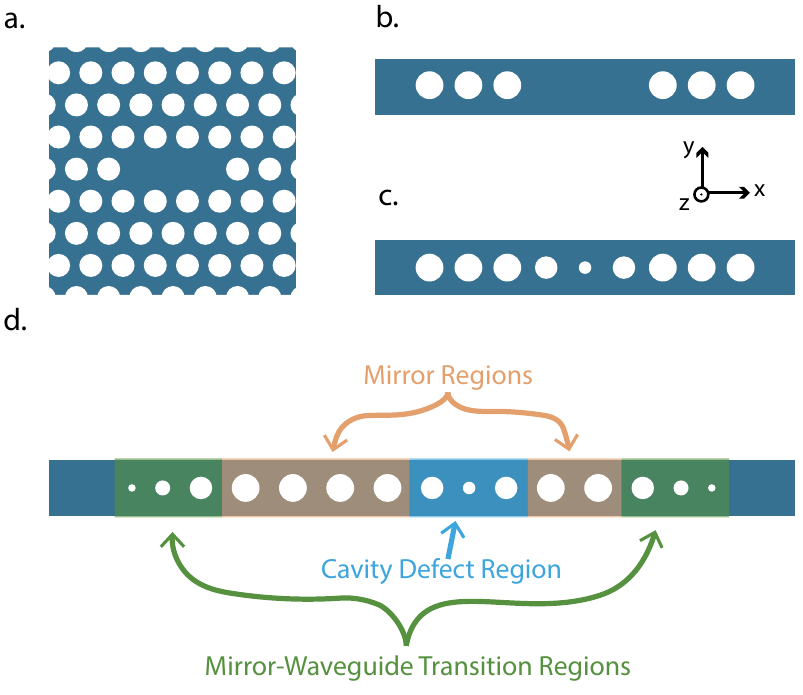}
    \caption{Photonic Cavity. (a) An L3 cavity in a 2-Dimensional photonic crystal exemplifies the original concept of a cavity formed by a defect in an otherwise perfect photonic crystal. (b) The same concept is used in 1-Dimensional photonic crystal cavities in nanobeams. Transverse confinement is provided by index guiding, and the cavity is formed by a defect in the photonic crystal. (c) Smoothly transitioning from the perfect crystal to the defect reduces scattering, increasing the quality factor of the cavity. (d) Modern nanophotonic cavities can be understood as being composed of three functionally distinct regions.}
    \label{fig:PCC}
\end{figure}

 The original concept of a photonic crystal cavity was realized as a defect within an otherwise perfect Bragg mirror. For example, so-called L3 cavities in air hole-based photonic crystals create a defect by replacing three adjacent air holes with completely filled dielectric (fig. \ref{fig:PCC}a) \cite{Joannopoulos2008}. The defect region supports optical modes that are forbidden elsewhere in the photonic crystal by a photonic band gap, thus a symmetric cavity is formed. In 1-D nanophotonic cavities (fig. \ref{fig:PCC}b), index guiding confines the light in y and z while the one-dimensional photonic crystal acts as a cavity along x, but the fundamental principle is identical to the L3 case. Most modern nanophotonic cavity designs exchange the abrupt introduction of a defect in favor of smoothly deforming the mirror region to create an extended defect that allows for gentle confinement and reduces scattering losses (fig. \ref{fig:PCC}c) \cite{Akahane2003,Quan2010a}. Additionally, to further reduce scattering in 1-D nanophotonic cavities, a region that transitions from the cavity mirrors to the coupling waveguide is typically appended to the cavity. The smooth defect and waveguide transitions are the essential elements of the modern nanophotonic cavity that offers a high Quality factor and wavelength scale mode volume V $\sim (\frac{\lambda}{n})^3$ in many material platforms. If an undercoupled or overcoupled cavity is desired, ``mirror" unit cells are added or removed on each side of the defect, allowing for directional coupling. Importantly, while this modern design differentiates the device into characteristic regions (mirror region, waveguide transition region, defect region) (fig. \ref{fig:PCC}d), it still relies on the original concept of a defect in an otherwise perfect Bragg mirror to create a cavity.
 
The nanophotonic cavities used in this experiment go beyond this ``three-region" design by allowing the mirror regions and their corresponding transition regions to be different on the left and right sides of the defect region. Figure \ref{fig:BS_Geom}c depicts a top-down view of this cavity. Since the mirror regions do not have identical unit cells, the defect region serves both as a localizing defect and a transition between the different mirror regions. Furthermore, the cavity is no longer simply a defect in an otherwise perfect Bragg mirror. Instead, each side of the cavity is a different Bragg mirror. This is similar to a Fabry–P\'erot cavity with different mirror strengths on either side, but the cavity mode is still determined by the supported modes of the defect regions is in traditional photonic crystal cavities. 

While our asymmetric cavity design does not correspond to the traditional picture of a defect in an otherwise perfect Bragg mirror, the additional design freedom proves useful for designing high cooperativity overcoupled nanophotonic cavities. Specifically, as we explain in the next section, we are able to create a mirror region that is effectively weaker for the cavity mode due to its relative detuning. Moreover, the quasipotential description we use to understand this unorthodox design reveals new insights into the limitations of the traditional approach.

We note that similar Purcell enhancement may be possible by operating in a slow-light nanophotonic waveguide coupled to a nanophotonic mirror instead of an overcoupled nanophotonic cavity. In fact, the enhancement theoretically diverges for 1-D systems as the group velocity goes to zero\cite{Lobet2020}. Preliminary calculations based on \cite{Lodahl2015} confirm that the enhancement could be comparable to within two orders of magnitude. Certain experimental challenges such as nanophotonic design, fabrication process development, directionality of emission, and other implementation details remain to be solved. Nevertheless, this offers potential for exciting future work.

\subsection{Quasipotential}

\begin{figure}
    \centering
    \includegraphics[width=\linewidth]{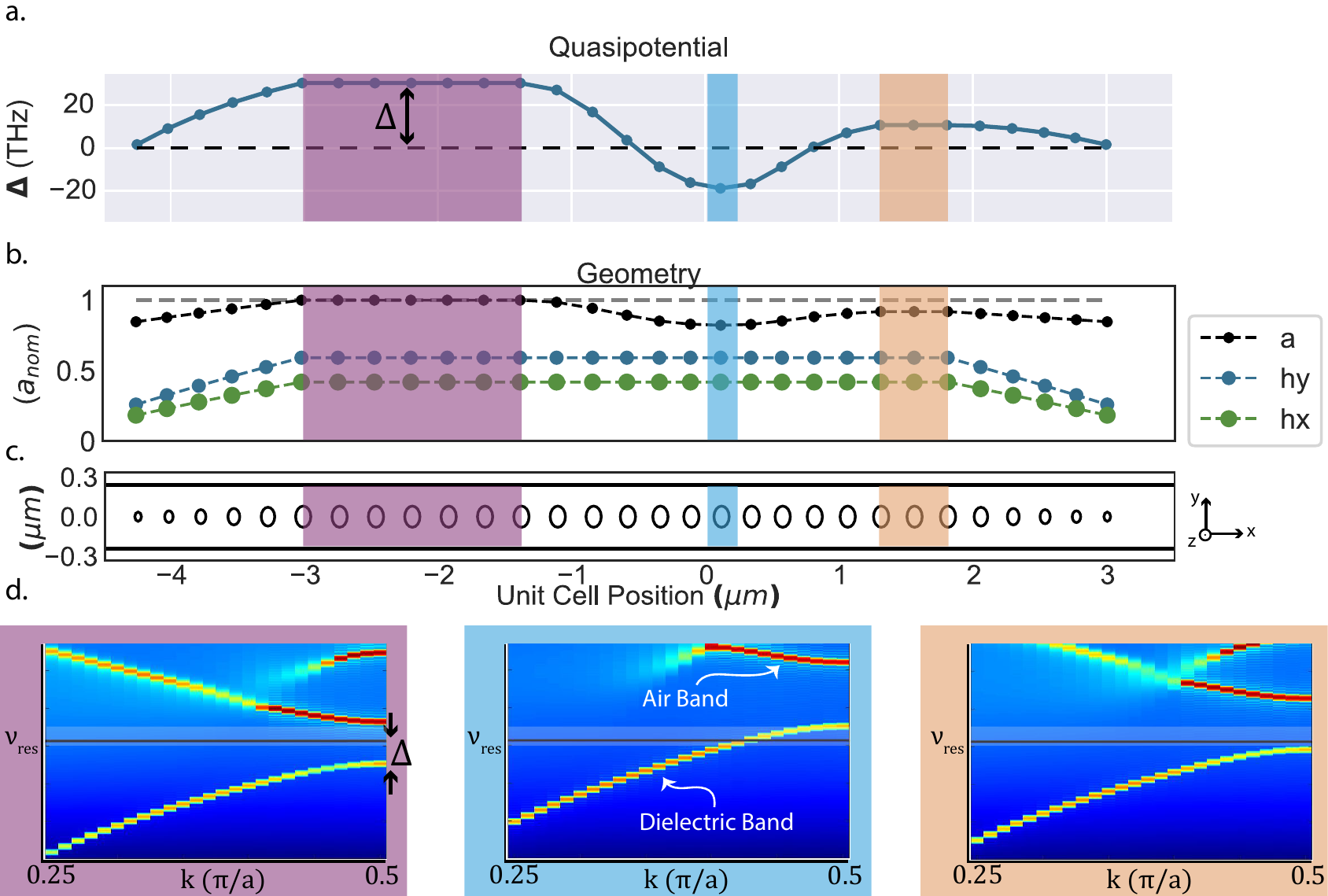}
    \caption{(a) Quasipotential for this device (b) Relative hole geometry parameters in units of the nominal lattice constant, $\text{a}_{\text{nom}}$. $\text{hy}_\text{nom}$ = 161 nm, $\text{hx}_\text{nom}$ = 114 nm, $\text{a}_\text{left}$ = 272 nm, $\text{a}_\text{right}$ = 250 nm. $\text{hy}_\text{nom}$ and $\text{hx}_\text{nom}$ correspond the cavity and mirror region holes' major and minor axes respectively. $\text{a}_\text{left}$ and $\text{a}_\text{right}$ are the lattice constants of the mirror regions on the left and right side of the cavity respectively. (c) Scale depiction of cavity top view. Waveguide width = 482 nm, triangle apex half-angle = $50^{\circ}$, maximum defect = 0.1392. (d) Simulated photonic bandstructure diagrams for representative unit cells from each cavity section. The color code indicates where the corresponding unit cell is located in the cavity. Quasipotential barrier height is defined as the detuning, $\Delta$, of the cavity resonance frequency (horizontal black lines) from the dielectric band edge at $\text{k} = 0.5$. Note the low potential barrier on the right side (orange) is due to the proximity of the dielectric band to the optical mode and \emph{not} a smaller band gap between the dielectric and air bands. Light white shaded regions represent the frequencies which would evanescently decay in both mirror regions.}
    \label{fig:BS_Geom}
\end{figure}

This asymmetric cavity is best understood by analogy to a quantum particle in a finite potential well. The idea is based on the similarity between a massive particle tunneling through a potential barrier and the evanescent decay of a photon in a photonic band gap \cite{Joannopoulos2008}. For our cavity design, the transmissivity of a mirror region is analogous to the tunneling probability through a potential barrier. The barrier height is given by the detuning of the mode from the dielectric band edge of each unit cell, and the barrier width is given by the number of unit cells in the mirror region. In this picture, important design targets like overcoupling and mode volume are made intuitive by the shape of the quasipotential. For example, the asymmetric quasipotential of the cavity used in these experiments is shown in fig. \ref{fig:BS_Geom}a, and it illustrates that the optimal cavity has both a lower and narrower barrier on the side which preferentially couples to the waveguide while maintaining a deep potential well necessary for the tight mode confinement characteristic of nanophotonic cavities. The simulated electric field overlay in fig. 2b of the main text illustrates this wavelength scale confinement (mode volume = 0.67 $\left(\frac{\lambda}{n}\right)^3$). 

On one side, we create an unusually weak mirror region to achieve good waveguide damping, while the other side has a strong mirror region with a high barrier to maintain low mode volume and effectively eliminate coupling to the unprobed waveguide. Using different unit cells in each mirror region is a departure from the traditional idea of a defect region contained within a uniform Bragg mirror \cite{Joannopoulos2008}, but it enables the cavity parameters to approach the best possible efficiency as defined by eqn. \ref{k_opt} and maintain the same small mode volumes seen in previous device designs.

We design the cavity to support a dielectric mode. (The electric field is concentrated in the diamond, not in the air). Therefore, we define the quasipotential barrier as the detuning of the dielectric band (lower band) from the resonance frequency of the cavity. Then, using Lumerical FDTD, we simulate the photonic band structure for each unit cell. We plot the detuning at each unit cell to reveal the quasipotential landscape of the nanophotonic cavity (fig. \ref{fig:BS_Geom}).

In fig. \ref{fig:BS_Geom}, we see the band structures corresponding to representative unit cells from the left mirror region, the defect region, and the right mirror region. Traversing the cavity from right to left, we see that the dielectric band starts very close to the optical mode. This small detuning corresponds to the low potential barrier visible in the orange section of fig. \ref{fig:BS_Geom}a, and it enables the tunneling that gives rise to the high coupling rate into the waveguide on this side of the cavity. Moving left to the blue indicated unit cell, the lattice constant decreases (fig. \ref{fig:BS_Geom}b), which shifts the bands to a higher frequency. This causes the dielectric band to overlap the resonant frequency of the cavity. Thus, the unit cells in the defect region support a propagating mode. Finally, we get to the purple, left mirror region. Here, the lattice constant has increased, shifting the bands down which results in the resonance frequency being far from the dielectric band near the center of the band gap. Thus, only evanescent modes are supported in this region, and they decay quickly. Similar to a particle in a finite well, the decay constant is proportional to the square root of the detuning.

Importantly, the detuning which weakens the mirror on the right side is only possible because the mirror region on the left side is significantly red detuned relative to the right mirror region, as illustrated by fig. \ref{fig:BS_Geom}d. Traditionally, a cavity's resonant frequency is near the center of the band gap of the mirror regions, as this is where the mirrors are strongest. In that case, the band structure for both mirror regions would be similar to the purple highlighted band structure in fig. \ref{fig:BS_Geom}d. On the other hand, in this cavity, modes that would have frequencies near the center of the band gap for the mirror region on the right side approach the air band for the mirrors on the left side. This is illustrated by the white shading in the three panels of fig. \ref{fig:BS_Geom}d. The upper edge of this shading is defined by the lowest air band in the mirror regions, that of the left side mirror cells, and the lower edge of this shading is defined by the highest dielectric band in the mirror regions, that of the right side mirror cells. Thus, any mode that could be formed by this combination of mirror cells would have to have a frequency in the white shading. In other words, the only modes which are allowed by the band structure of the left side force the mirrors of the right side to be weaker. This fine-tuning of the barrier height by tuning the frequency of the bands relative to the resonant mode is a novel and useful feature of this asymmetric design approach.

While the quasipotential is a useful tool for understanding an optimized design, it's important to note that it is a simplification of the nanophotonic cavity that neglects important degrees of freedom, namely, it does not capture information about the transverse mode profile. Therefore, a cavity should not be designed solely by concatenating unit cells to create a desired quasipotential. Care should also be taken to ensure that the transverse mode profile of each subsequent unit cell does not differ significantly from its neighbors as this would induce significant scattering, reducing the cavity quality factor. In our cavity design, we ensure this smooth changing mode profile by smoothly changing the unit cell parameters.

\subsection{Mode Volume Trade-off}

\begin{figure}
    \centering
    \includegraphics[width=\linewidth]{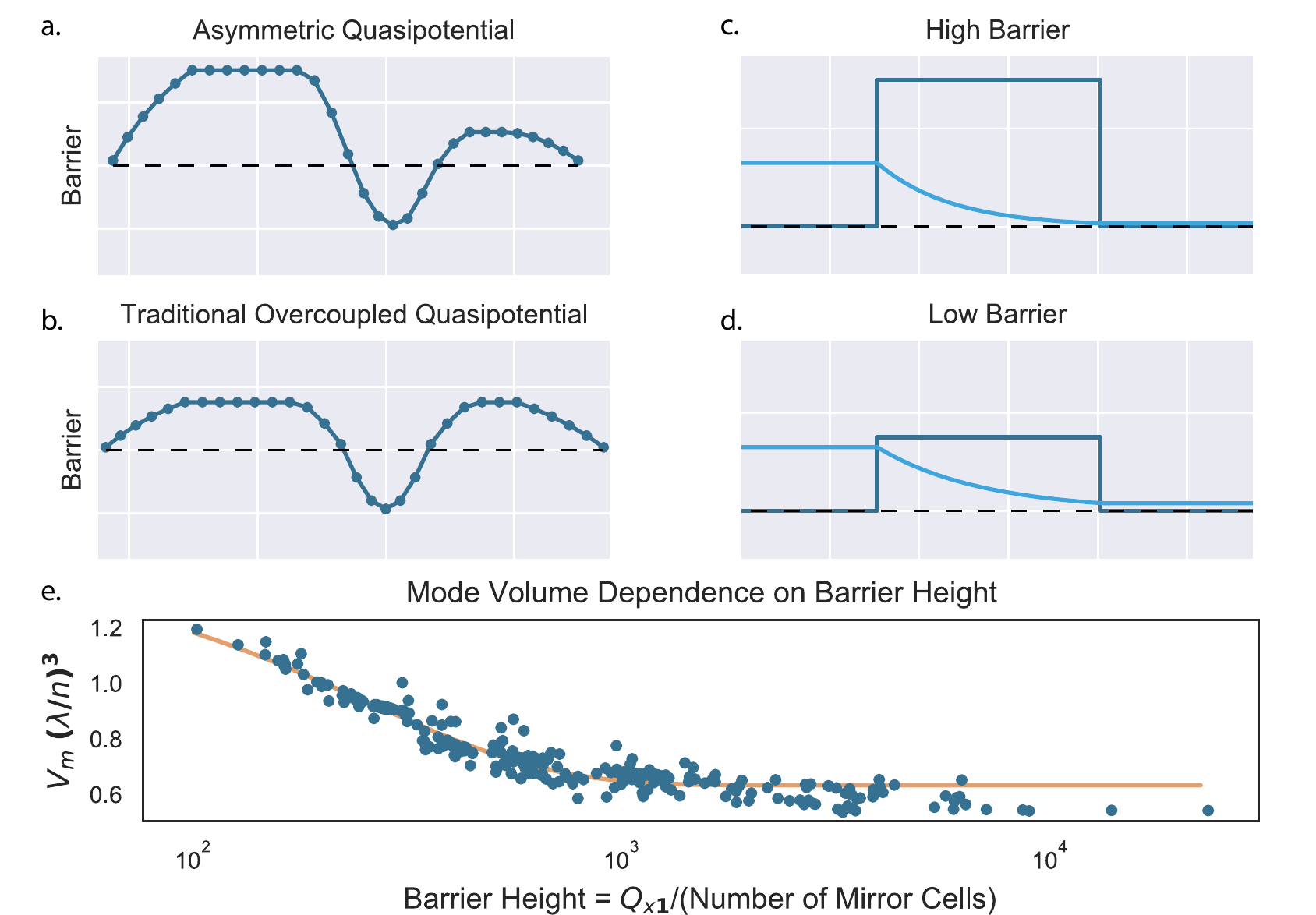}
    \caption{Mode volume increases for lower potential barriers due to slower decay within the mirror region. (a) The asymmetric quasipotential of the cavities used in this experiment takes advantage of both the tight confinement of a high barrier on the left-hand side for low mode volume and the fast coupling of a low barrier on the right-hand side for good overcoupling. (b) In order to make an overcoupled cavity with identical unit cells in each mirror region, the barrier must be low, to allow for good waveguide coupling, but long on one side to enable asymmetric coupling. This restrictive design creates a trade-off between overcoupling and mode volume because a high barrier (c) strongly suppresses the evanescent field, while a low barrier (d) suppresses the field more weakly. (e) Simulation data of many cavities accumulated during a geometry optimization sweep illustrate that low barrier heights correspond to larger mode volumes.}
    \label{fig:Vmode}
\end{figure}

In a highly overcoupled cavity, the coupling rate to one side is much higher than the coupling rate to the other side and to free space. In traditional nanophotonic cavities, this coupling rate difference is achieved simply by putting a different number of mirror unit cells on each side of the defect region. This approach was used successfully in \cite{Tiecke2014}. However, there is a minimum number of mirror unit cells below which the cavity suffers increased scattering losses. (In our system, the minimum number is typically three unit cells.) Therefore, if the directional coupling rate is too low at the minimum barrier width, the barrier height of the mirror unit cells needs to be lowered. Lowering the barriers increases the mode volume of the cavity, as illustrated in fig. \ref{fig:Vmode}e.

Atom-Cavity cooperativity depends on both the cavity quality factor Q and the mode volume V as given by the Purcell enhancement $\frac{Q}{V}$. Thus, in order to maximize cooperativity, one must maximize the cavity quality factor and minimize the mode volume. Conversely, for a given quality factor, a cavity with a larger mode volume will have lower cooperativity, and, in this case, a lower single-photon generation efficiency.

In a cavity with the same height barrier on each side of the defect region, there is an intrinsic trade-off between overcoupling and mode volume. This trade-off is intuitively obvious from the fact that the field decay is stronger in a high barrier (fig. \ref{fig:Vmode}c) than in a low barrier (fig. \ref{fig:Vmode}d). Thus, when using lower barriers to allow for better waveguide coupling, mode confinement is worse, causing an increase in mode volume.

The quasipotential of this cavity (fig. \ref{fig:Vmode}a) compared with that of a traditional overcoupled cavity (fig. \ref{fig:Vmode}b) shows that allowing the cavity to have different barrier heights on each side of the defect enables both small mode volume (due to the strong mode confinement of the high barrier) and good overcoupling (due to the low barrier on the coupling side). On the other hand, the traditionally designed cavity must sacrifice confinement on the strong mirror side in order to enable good coupling on the weak mirror side. 

In fig. \ref{fig:Vmode}e, simulation data from cavities from several optimization runs illustrate the increasing mode volume for lower barriers. Each of the data points represents a single cavity with seven mirror unit cells on the left side of the cavity and between one and five mirror unit cells on the right side of the cavity. Given that the field decay rate is determined by the total barrier area, we define an effective barrier height as the directional quality factor $Q_{x1}$ (dimensionless inverse coupling rate to the right) divided by the number of mirror unit cells.

\subsection{Fabrication and Design Details}
The nanophotonic cavities used for this work were fabricated with the methods described in \cite{Nguyen2019a}. One notable subtlety is that the fabrication procedure of a new cavity design must be tuned up for optimal resist thickness. The resist has to be thin enough to allow the holes to completely clear during the top-down etch but thick enough that the holes are protected during the angled etch. Historically, we found that fabrication of cavities with wildly varying hole sizes was challenging because the appropriate resist thickness for holes of a particular size would not work for holes of different size in the cavity. With this in mind, we restricted our design space to maintain identical hole dimensions throughout the defect and mirror regions. Then, the defect and mirror asymmetry were facilitated by a changing lattice constant.

\begin{figure}
    \centering
    \includegraphics[width=\linewidth-\linewidth/4]{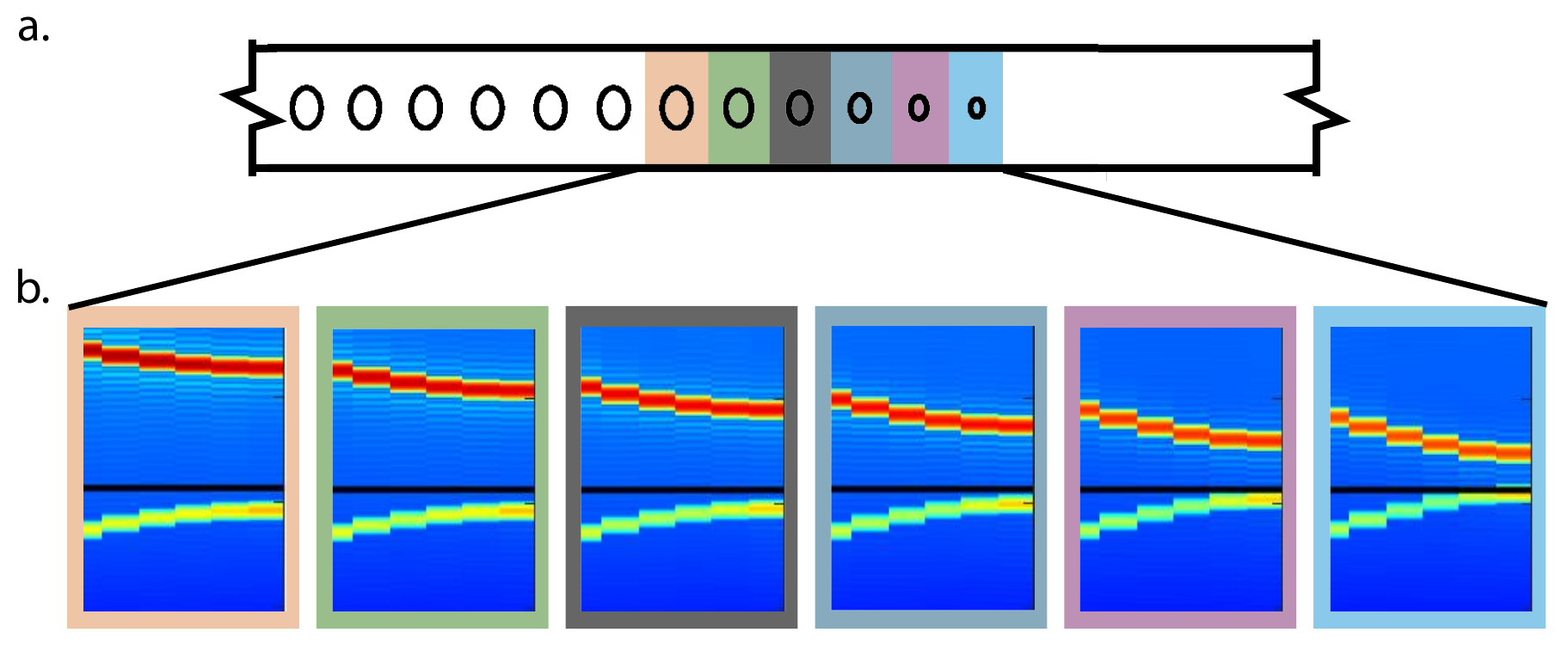}
    \caption{The mirror-waveguide transition region for the right-hand side of the device (a) and each unit cell's corresponding simulated band structure (b). The frequency of the resonant cavity mode is indicated by the horizontal black line. Resonant light exiting the cavity is pushed into the dielectric band so that it can smoothly transition to the waveguide mode as the photonic band gap is closed.}
    \label{fig:BS_taper}
\end{figure}

One may note that in the mirror-waveguide transition regions, we relax the hole dimension restriction. Here we have both a linearly decreasing lattice constant and linearly decreasing hole dimensions as illustrated in \cref{fig:BS_Geom}b and \cref{fig:BS_taper}. This is necessary to facilitate low-loss transitions between the waveguide and the cavity. The final lattice constant before the waveguide is chosen to match the periodicity of the standing wave formed by the incoming and reflected light as set by the effective refractive index of the waveguide and the optical frequency of the mode. In choosing this lattice constant, we aim to have the electric field nodes centered on the holes, so that the incident electric field is initially minimum when encountering an interface, thereby reducing scattering. The complementary view from the perspective of the outgoing cavity light is that the smooth reduction of these holes and the decreasing lattice constant effectively pushes the mode back into the dielectric band before closing the gap entirely.

The varied hole size does result in more significant fabrication error. Indeed, the outer most holes of the taper often do not come out in the fabricated structures. These fabrication errors could be mitigated through improved Proximity Effect Correction or manual adjustments and iterative fabrication. Nevertheless, we empirically find that the presence of an imperfect transition region in the fabricated structures reduces scattering as compared with having no transition region. Because the field intensity is much lower in these regions, the performance of the device is less sensitive to fabrication errors there.

As in \cite{Nguyen2019a}, the cavity design is optimized using gradient ascent. However, the figure of merit in this case is more complicated than simple Purcell enhancement so that cavity fitness additionally accounts for the cavity overcoupling. Therefore, the figure of merit used in the optimization which produced this experiment's design is $$\text{Fitness} = \sqrt{\frac{Q_{\text{scat}}}{Q_{\text{wvg}}}*\frac{Q_{\text{left}}}{Q_{\text{right}}}*\frac{Q*Q_{\text{scat}}}{V^2}*e^{-(5^{-2}(\lambda_0-\lambda)^2)}}$$
Here, $\frac{Q_{\text{scat}}}{Q_{\text{wvg}}}$ captures the relative waveguide damping rate versus scattering loss rate, with $Q_{\text{wvg}} = (Q_{\text{left}}^{-1}+Q_{\text{right}}^{-1})^{-1}$. $\frac{Q_{\text{left}}}{Q_{\text{right}}}$ captures how directional the waveguide damping is. $\frac{Q*Q_{\text{scat}}}{V^2}$, with $Q = (Q_{\text{wvg}}^{-1}+Q_{\text{scat}}^{-1})^{-1}$ is effectively Purcell enhancement but weighted to value improvements in scattering quality factor as well as total quality factor. Note that $Q \equiv \frac{\nu_\text{res}}{\kappa}$ with $\nu_\text{res}$ the resonant frequency of the cavity, which is about 407 THz here. Finally, $e^{-(5^{-2}(\lambda_0-\lambda)^2)}$ is a smooth penalty that restricts the optimizer to find results near the target resonance wavelength. Although ad hoc, it was necessary to add the Gaussian resonance penalty to constrain the optimizer to produce useful results. Without the resonance penalty, the cavity mode volumes would artificially be reduced to zero because the optimizer would create cavities with resonances outside of the source bandwidth. Similar Gaussian penalties may be promising in future work for targeting specific cavity parameters, for example, a specific value of $Q_{\text{right}}$.

The cavity geometry parameters varied during optimization were the lattice constant on the left side (aL), the lattice constant on the right side (aR), the hole height in the mirror regions (hy), the hole width in the mirror region (hx), and the maximum defect (max$\_$def). The functional form of the defect is identical to that described in \cite{Nguyen2019a} except we multiply the defect by a linear slope term to make a smooth transition between the different lattice constants on each side of the cavity.

The design and optimization process is computationally very expensive. Each full 3D cavity simulation is performed in Lumerical FDTD with a manual 12 nm resolution mesh region around the whole cavity as well as a high auto mesh accuracy = 5 for the remainder of the cavity simulation region. Simulations were performed using the Harvard University Research Computing cluster ODYSSEY (now known as CANNON). Though the gradient ascent proceeds serially, the time for each cavity simulation step is greatly reduced by exploiting the FDTD MPI. Typical optimizations are performed with 64 GB of RAM and up to 50 cores, reducing individual cavity simulation times from $\sim20$ minutes on a typical laptop to $\sim3$ minutes. 

\subsection{Cavity Fit / Overcoupling}

\subsubsection{Evidence of Overcoupling}
\label{sec:overc}

\begin{figure}
    \centering
    \includegraphics[width=\linewidth]{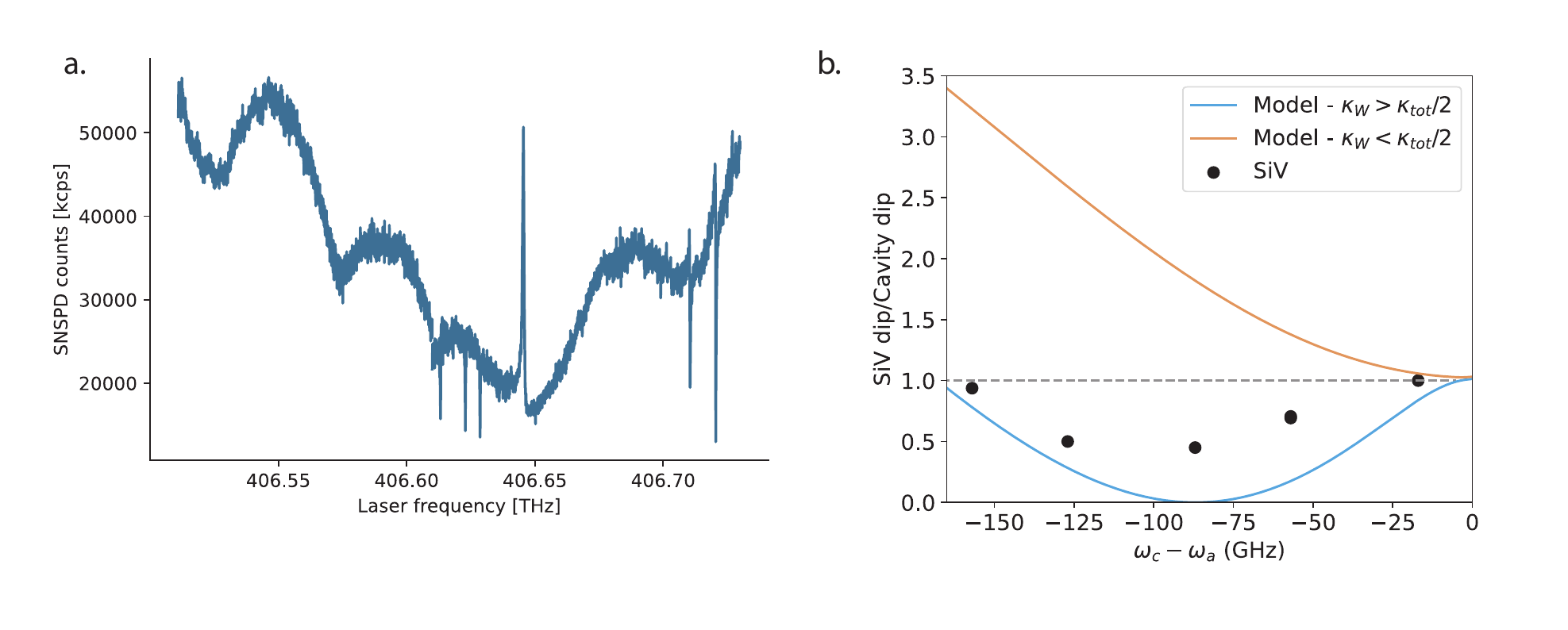}
    \caption{(a) Reflection spectrum of the nanophotonic cavity coupled to six SiV centers. The fact that some SiV features dip below the cavity minimum is a signature feature of an overcoupled cavity. The slow modulation of the cavity reflection is due to interference created by reflections within the fiber network. (b) Measured minimum reflection of the SiV spectral feature normalized to the cavity dip versus its detuning from the cavity resonance. Dips below the cavity minimum (grey dashed line) are again observed, demonstrating overcoupling of a cavity. For comparison, theoretical SiV to cavity dip ratios, as obtained by fitting \protect\cref{eq:ref}, for an equivalent overcoupled (blue) or undercoupled (orange) cavity are shown.}
    \label{fig:cavity_scan}
\end{figure}

In CQED, whether the cavity is undercoupled, critically coupled, or overcoupled qualitatively changes how the cavity interacts with incident light and thus how a coupled emitter can modify the cavity's spectrum. In this experiment, we probe the cavity in reflection, monitoring the intensity of the light that returns from the cavity. Of the three regimes, a critically coupled cavity is easiest to distinguish because it exhibits a characteristic full contrast reflection dip, a Lorentzian feature that goes from the cavity reflection maximum when the probe light is off-resonant with the cavity to approximately zero photons reflected when the probe light is resonant with the cavity. This dip results from the destructive interference between the light directly reflected from the front mirror of the cavity and the reemitted cavity field. In reflection, a bare overcoupled and bare undercoupled cavity can have the same intensity spectrum, a Lorentzian dip that does not go to zero. However, the cause of the partial contrast dip is different for these complementary regimes. In the undercoupled cavity, the reflection dip does not go all the way to zero because the reemitted cavity field is always \emph{weaker} than the directly reflected light. Thus, perfect destructive interference can never occur. On the other hand, for an overcoupled cavity, the reemitted cavity field always has a \emph{larger} amplitude than the directly reflected light, and thus perfect destructive interference can never occur. Importantly, these distinct mechanisms result in different reflection spectra for overcoupled and undercoupled cavities coupled to emitters. Namely, \textbf{only emitters in overcoupled cavities can produce spectral features that dip below the cavity minimum. }

It is instructive to consider the electric field of resonant light reflected from a cavity. The reflected field of the bare cavity is $E_{\text{refl}} = 1-\frac{2\kappa_{\text{w}}}{\kappa_{\text{w}}+\kappa_{\text{s}}}$. Thus, when $\kappa_{\text{s}} \not = \kappa_{\text{w}}$, there is not a full contrast dip. Putting an emitter in the cavity introduces a new loss channel which is $g^2/\gamma$ when the emitter is resonant with the cavity. Then, the reflected field is $E_{\text{refl}} = 1-\frac{2\kappa_{\text{w}}}{\kappa_{\text{w}}+\kappa_{\text{s}}+g^2/\gamma}$, and full contrast is possible when the cavity is overcoupled ($\kappa_{\text{w}} > \kappa_{\text{s}}$).

\Cref{fig:cavity_scan}a shows a reflection spectrum of the nanophotonic cavity used for this experiment. Sharp reflection dips due to SiVs which dip below the cavity minimum are clearly observed. Thus, we conclude that this cavity is, in fact, overcoupled. We note that there is a slow modulation in the cavity spectrum which we attribute to interference from stray reflections in the fiber network. However, these interferences are much broader than the SiV features we observe in the reflection spectrum and the photons generated from the SiV ($\sim50$ GHz vs $10$-$100$'s MHz) and thus do not significantly impact the experiment.

To provide further evidence of cavity overcoupling, the SiV's minimum reflection dip was measured as a function of its detuning from the cavity for a different device (\Cref{fig:cavity_scan}b). The detuning was varied via gas tuning of the cavity \cite{Nguyen2019a}. In addition to observing the SiV dip below the cavity dip, we observe that the SiV reflection dip is minimized at a finite detuning away from the cavity. This behavior is also indicative of an overcoupled cavity (see calculated model curves in \Cref{fig:cavity_scan}b). We note the discrepancy between the theoretical overcoupled cavity dip versus detuning and the measured values is due to a combination of spectral diffusion, stray reflections, and dark counts limiting the minimum reflection observable in our system.

\subsubsection{Estimation of Cooperativity}

\begin{figure}
    \centering
\includegraphics[width=0.65\textwidth]{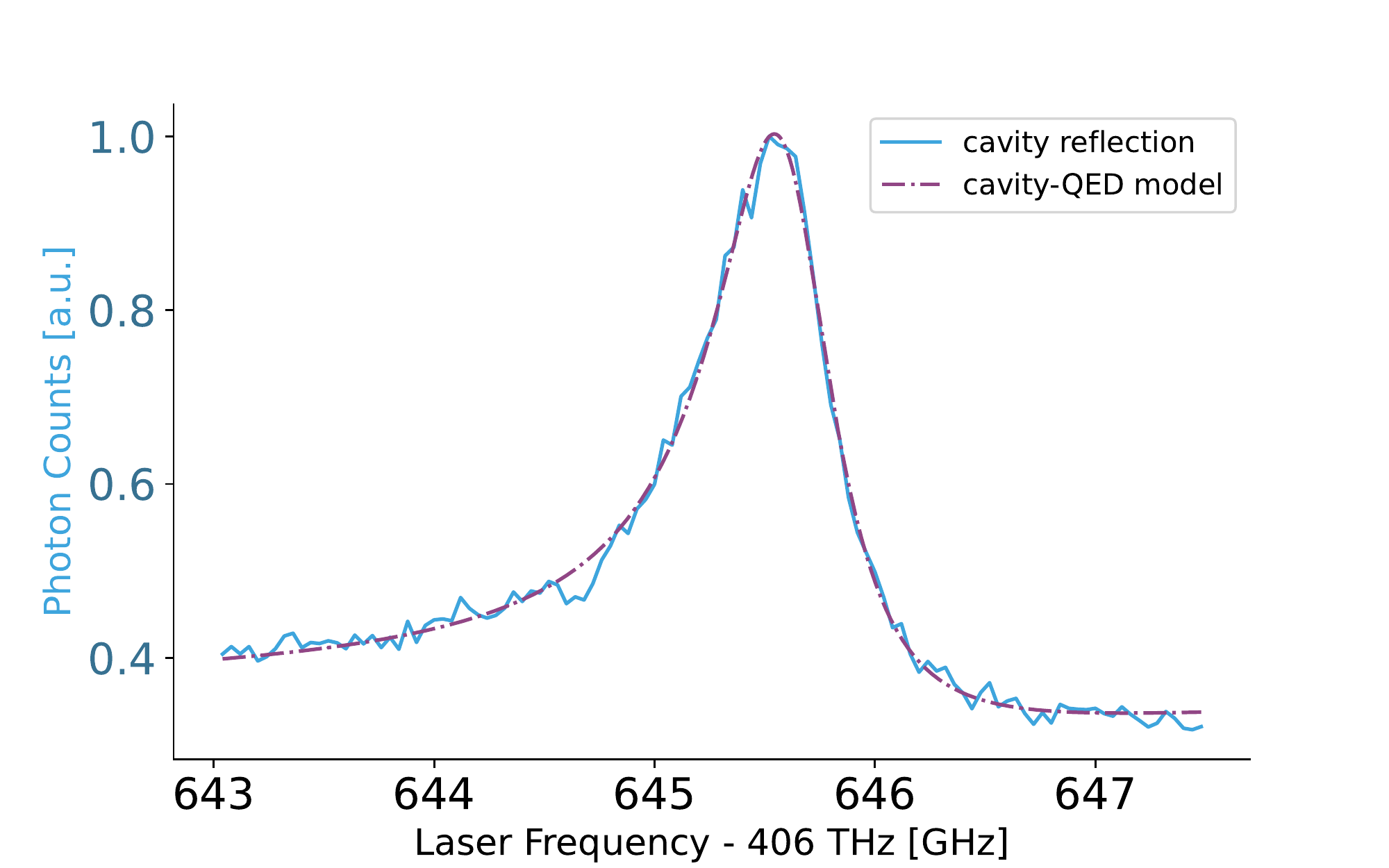}
    \caption{Reflection spectrum of nanophotonic cavity coupled to a single SiV center and fit to \protect\cref{eq:ref}.}
    \label{fig:reflection}
\end{figure}

The total cavity damping rate $\kappa_{\text{tot}}$ along with the waveguide damping rate $\kappa_{w}$ were determined by fitting a broad cavity reflection spectrum taken with a spectrometer (Horiba iHR-550). For this fit, we assume the cavity is overcoupled (see section \ref{sec:overc}). The single photon Rabi rate $g$ was determined by fitting the reflection spectrum of the atom-cavity system near the frequency of the SiV's to the following intensity reflection coefficient expression. This expression is derived using input-output formalism applied to our Jaynes-Cummings Hamiltonian \cite{RevModPhys.87.1379}:

\begin{align}
\mathcal{R}(\omega) = {\left | 1 - \frac{2 \kappa_w}{2i(\omega - \omega_c) + \kappa_{\text{tot}} + 4g^2 / (2i (\omega - \omega_a) + \gamma)} \right |}^2
\label{eq:ref}
\end{align}
Here, $\omega_C$ is the resonance frequency of the nanophotonic cavity, $\omega_a$ is the SiV resonance frequency, and $\gamma$ is the linewidth of the SiV (FWHM). The complete set of cavity-QED parameters

\begin{align}
\{ g , \kappa , \gamma \} = 2 \pi \times \{\SI{6.81 \pm 0.06}{\giga\hertz}, \SI{328.74 \pm 3.43}{\giga\hertz}, \SI{100}{\mega\hertz} \}
\label{eq:fitparams}
\end{align}
gives rise to the fit shown in \cref{fig:reflection} and results in cooperativity $C = \frac{4g^2}{\kappa\gamma} = 5.64 \pm 0.12$.\\

\section{Theoretical Description of Single-Photon Generation}

To gain intuition for single-photon generation by slow optical pumping, we model the SiV as a $\Lambda$-system coupled to a continuum of modes, neglecting the coupling to the nanophotonic cavity for now. In the rotating frame of the laser frequency $\nu$ of the control pulse, and by making the rotating frame approximation, we can express the system Hamiltonian as

\begin{equation}
    H = -\Delta \ket{\downarrow^\prime, 0}\bra{\downarrow^\prime, 0} -  \Omega\ket{\downarrow^\prime, 0}\bra{\uparrow, 0} + \sum_k \left[ \omega_k \ket{\downarrow, 1_k}\bra{\downarrow, 1_k} - g_k \ket{\downarrow^\prime, 0}\bra{\downarrow, 1_k} \right] + \text{h.c}
\end{equation}

Here, $\Delta$ is the detuning of the control pulse from the $\ket{\uparrow} \rightarrow \ket{\downarrow^\prime}$ transition, $k$ indexes the environmental photon modes with energies $\omega_k$ (in the rotating basis) expressed as $\omega_k = \nu_k - \nu - \omega_{\uparrow \downarrow}$, with $\nu_k$ being the frequency of the environmental photon and $\omega_{\uparrow \downarrow}$ being the energy splitting between the $\ket{\downarrow} \rightarrow \ket{\uparrow}$ transition, and $g_k$ is the single photon Rabi frequency.\\

Applying the Wigner-Weisskopf approximation for spontaneous emission \cite{weisskopf1997berechnung} and writing out the equations of motion for the diagonal density-matrix element $\rho_{\downarrow^\prime, \downarrow^\prime} = \ket{\downarrow^\prime, 0}\bra{\downarrow^\prime, 0}$, $\rho_{\uparrow, \uparrow} = \ket{\uparrow, 0}\bra{\uparrow, 0}$, we get\\ 

\begin{align} 
\dot{\rho}_{\uparrow, \uparrow} &= i \Omega^*(t) \rho_{\downarrow^\prime, \downarrow^\prime}\\
\dot{\rho}_{\downarrow^\prime, \downarrow^\prime} &= i \Delta \rho_{\downarrow^\prime, \downarrow^\prime} + i \Omega(t) \rho_{\uparrow, \uparrow} - \gamma \rho_{\downarrow^\prime, \downarrow^\prime} \\
\end{align}
With the spontaneous emission rate
\begin{equation}
    \gamma = 2 \frac{\omega_{13}^3 |\mu_{\downarrow \rightarrow \downarrow^\prime}|^ 2}{3 \pi \epsilon_0 \hbar c^3}
\end{equation}
Here, $\mu_{\downarrow \rightarrow \downarrow^\prime}$ is the transition dipole element for the $\ket{\downarrow} \rightarrow \ket{\downarrow^\prime}$ transition. We now assume that we drive the $\ket{\uparrow} \rightarrow \ket{\downarrow^\prime}$ transition on resonance ($\Delta = 0$) and get 
\begin{equation}
\dot{\rho}_{\downarrow^\prime, \downarrow^\prime} = i \Omega(t) \rho_{\uparrow, \uparrow} - \gamma \rho_{\downarrow^\prime, \downarrow^\prime}
\end{equation}
If we now assume a time-independent Rabi drive $\Omega(t) = \Omega$ with $\Omega \ll \gamma$ , we can adiabatically eliminate the time dynamics of the exited state $\ket{\downarrow^\prime, 0}$ and can write

\begin{equation}
   \rho_{\downarrow^\prime, \downarrow^\prime} = \frac{i \Omega}{\gamma} \rho_{\uparrow, \uparrow}
\end{equation}
It should be noted that the adiabaticity condition can easily be fulfilled, since $\Omega \propto \mu_{\uparrow \rightarrow \downarrow^\prime}$, and the cross transition $\ket{\uparrow} \rightarrow \ket{\downarrow^\prime}$ is much weaker than the spin conserving transition $\ket{\downarrow} \rightarrow \ket{\downarrow^\prime}$: $ |\mu_{\downarrow \rightarrow \downarrow^\prime}|^ 2 \gg |\mu_{\uparrow \rightarrow \downarrow^\prime}|^ 2$.

Assuming $\rho_{\uparrow, \uparrow}(t=0) = 1$, one can now define a modified fluorescence rate $\Gamma_{\text{fl}}$ as 

\begin{equation}
    \Gamma_{\text{fl}} = \gamma |\rho_{\downarrow^\prime, \downarrow^\prime}|^2 =  \gamma \left(\frac{|\Omega|^2}{\gamma^2}\right) \label{gammafl} \ll \gamma
\end{equation}
The photon emission now occurs at timescales dictated by $\Gamma_{\text{fl}}$, which are much slower than the spontaneous emission rate $\gamma$. Notably, the timescales are only limited by the coherence properties of the $\ket{\uparrow}, \ket{\downarrow}$ qubit levels and the noise of the drive laser.\\

So far, this treatment does not explicitly take the coupling between the SiV center and the nanophotonic cavity into account. Assuming a cavity decay rate $\kappa$, the Purcell regime \cite{RevModPhys.87.1379} is characterized by: 

\begin{align}
\label{eq:purcellregime}
  \kappa \gg g \gg \gamma
\end{align}

In this regime, the linewidth of an SiV close to the cavity's resonance frequency broadens and becomes $\Gamma = (C+1) \gamma$, with $C =4 \frac{g^2}{\kappa\gamma}$. Working with relatively broad cavities ($\kappa \approx 2 \pi \times 300$ GHz) facilitates the satisfaction of \cref{eq:purcellregime} and motivates borrowing the intuition for scattering into a continuum of modes. 
For $C>1$, the adiabaticity condition is thus relaxed to $\Omega \ll \Gamma$. Using our simulation results for the Gaussian photon, we can independently verify that the adiabaticity condition is met, see \cref{fig:simulations} (b).

\section{Density Matrix Simulations of Photon Generation}
\subsection{Model Setup}
\label{sec:density}
In the rotating frame of the frequency of the control pulse, we can express our Hamiltonian in the joint atom-cavity basis:

\begin{equation}\label{eq:Hs}
\mathbf{H}_{\text{S}} = \Delta \ket{\downarrow^\prime, 0} \bra{\downarrow^\prime, 0} + (\Delta-\Delta_{C})\ket{\downarrow, 1}\bra{ \downarrow, 1} + \Omega(t)\ket{\downarrow^\prime, 0}\bra{ \uparrow, 0} + g\ket{\downarrow, 1}\bra{\downarrow^\prime, 0} + \text{h.c.}
\end{equation}
Here, $\Delta$ is the detuning of the control pulse from the $\ket{\uparrow} \rightarrow \ket{\downarrow^\prime}$ transition, $\Delta_{C}$ is the detuning between the $\ket{\uparrow}\rightarrow \ket{\downarrow^\prime}$ transition and the cavity resonance, $\Omega(t)$ is the time-dependent Rabi drive induced by the control pulse, $g$ is the coupling rate to the cavity mode, and $\hbar$ is set to 1. \\

\begin{figure}
    \centering
\includegraphics{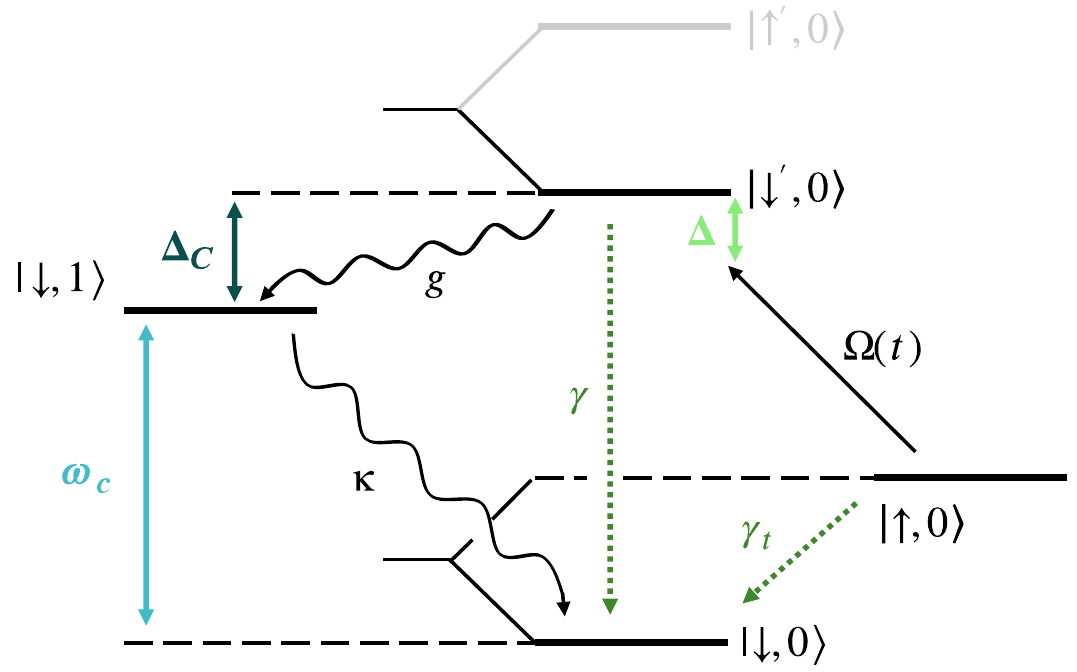}
    \caption{Level diagram of our system whose dynamics are encoded in Eq. \protect\ref{eq:Hs}. The Zeeman interaction lifts the degenracy of the $\ket{\uparrow}$ and $\ket{\downarrow}$ ($\ket{\uparrow}^\prime$ and $\ket{\downarrow}^\prime$) states.}
    \label{fig:levels}
\end{figure}

The following Lindblad collapse operators are used to describe the interaction between the system and the environment:

\begin{equation}\label{eq:c_ops}
\mathbf{L_1} = \sqrt{\kappa}\ket{\downarrow, 0}\bra{\downarrow, 1},
\mathbf{L_2} = \sqrt{\gamma}\ket{\downarrow, 0}\bra{\downarrow^\prime, 0},
\mathbf{L_3} = \sqrt{\gamma_{\text{t}}} \ket{\downarrow, 0}\bra{\uparrow, 0}
\end{equation}
Here, $\kappa$ is the rate at which the photon leaks out of the cavity. For the purpose of this model, we assume that we collect all photons emitted from the cavity, such that it is not necessary to distinguish between the cavity decay rate $\kappa$ and the waveguide decay rate $\kappa_w$. $\gamma$ is spontaneous emission rate for the optical $\ket{\downarrow^\prime}\rightarrow \ket{\downarrow}$ transition, and $\gamma_{\text{t}}$ is the decay rate within the spin-qubit. The level diagram for this system is shown in fig. \ref{fig:levels}.\\

We can now use the Lindblad-form of the Master equation to describe the time-dynamics of the the density operator of the system:

\begin{equation}\label{eq:master-eq}
    \mathbf{\dot \rho} = -i[ \mathbf{H}_{\text{S}}, \mathbf{\rho}] + \sum_i (\mathbf{L_i}\mathbf{\rho}\mathbf{L_i^{\dagger}} - \frac{1}{2}\mathbf{L_i^\dagger}\mathbf{L_i}\mathbf{\rho}-\frac{1}{2}\mathbf{\rho}\mathbf{L_i^\dagger}\mathbf{L_i})
\end{equation}

This master equation can be numerically evaluated using Python's \texttt{QuTip} package \cite{qutip2}, an open-source software for simulating the dynamics of open quantum systems.

\subsection{Simulation Results: Photon Pulseshape}

We encode our effective Hamiltonian (Eq. \ref{eq:Hs}) and collapse operators (Eq. \ref{eq:c_ops}) with the appropriate parameters and simulate how the system evolves when being driven by an arbitrary time-dependent Rabi pulse $\Omega(t)$. For the example of the Gaussian photon depicted in the main text, fig. 3 (b), we can parameterize the Rabi drive as a Gaussian pulse:

\begin{equation}\label{eq:rabidrive}
\Omega(t) = \Omega_{0} \exp{\left [ - \frac{1}{2} {\left( \frac{t - \mu}{\sigma} \right) }^2 \right ]}
\end{equation}
Using this expression for $\Omega(t)$, we can use the \texttt{mesolve} function in \texttt{QuTip} to solve the master equation and obtain the state of the system $\rho$ at each given time step. To extract the intensity profile of the emitted photon, we then find $\sqrt{\kappa}\braket{\downarrow, 1|\rho|\downarrow, 1}$, where $\braket{\downarrow, 1|\rho|\downarrow, 1}$ is the population of the cavity $\ket{\downarrow, 1}$.\\

\begin{figure}
    \centering
    \includegraphics[width=0.99\textwidth]{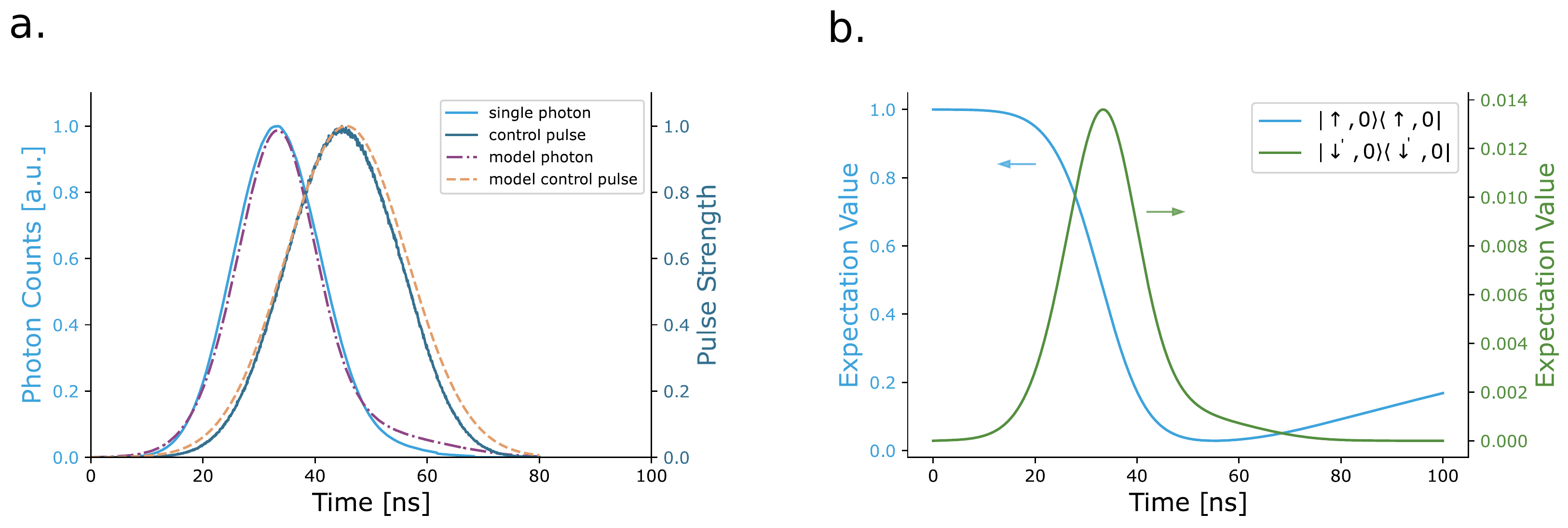}
    \caption{(a) Model fit of a Gaussian photon generated from the given pulse, including photon and pulse data also shown in main text fig. 3 (b) Simulated expectation values of operators for the same simulation as in (a). As population is fully transferred out of the initial spin state (blue curve), the population of the excited state (green curve) remains low. Arrows indicate relevant y-axis.}
    \label{fig:simulations}
\end{figure}

This numerical model enables us to design pulse shapes to produce a desired photon waveform. By optimizing over pulse parameters to fit a target photon shape, we can use our dynamical model of the atom-cavity system to predict pulses that will generate a specific photon. We can also use our model to replicate experimentally obtained photon and pulse data, as shown in fig. \ref{fig:simulations} (a). This is accomplished by first fitting photon data to obtain a predicted pulse shape, then inputting the experimental pulse data into our model to generate a photon, and finally averaging the two results. Table \ref{tab:tablesim} shows the numerical values of the CQED parameters for the simulation depicted in fig. \ref{fig:simulations}.\\

Figure \ref{fig:simulations} (b) confirms through the results of a density matrix simulation with realistic parameters that the excited state is only slightly populated during the photon generation sequence as expected. The dominance of $\kappa$ and $g$ over the spontaneous decay rate $\gamma$ allows coherent population transfer from the spin ground state to the single photon without suffering significant decoherence from $\gamma$. \\

\begin{table*}
\begin{ruledtabular}
\begin{tabular}{ccccc}
 Variable&Description&Value
\\ \hline
 $\Delta_C / 2 \pi$&SiV - cavity detuning&19.88  GHz \\
  $g / 2 \pi$&Single photon Rabi frequency&6.81 GHz \\
   $\Omega_0/ 2 \pi$& Rabi drive peak amplitude &194 MHz\\
 $\kappa / 2 \pi $&FWHM of nanophotonic cavity &329 GHz \\
  $\gamma / 2 \pi $&Spontaneous emission rate for $\ket{\uparrow} \rightarrow \ket{\downarrow^\prime}$ transition &100 MHz \\
  $\gamma_{\text{t}} / 2 \pi$& Decay rate of spin qubit &50 kHz\\

\end{tabular}
\end{ruledtabular}
\caption{ Numerical vales of model parameters used in simulations results shown in \protect\cref{fig:simulations}}
\label{tab:tablesim}
\end{table*}

\subsection{Simulation Results for Correlation Functions}
\label{section:g2sim}

\begin{figure}
    \centering
    \includegraphics[width=0.99\textwidth]{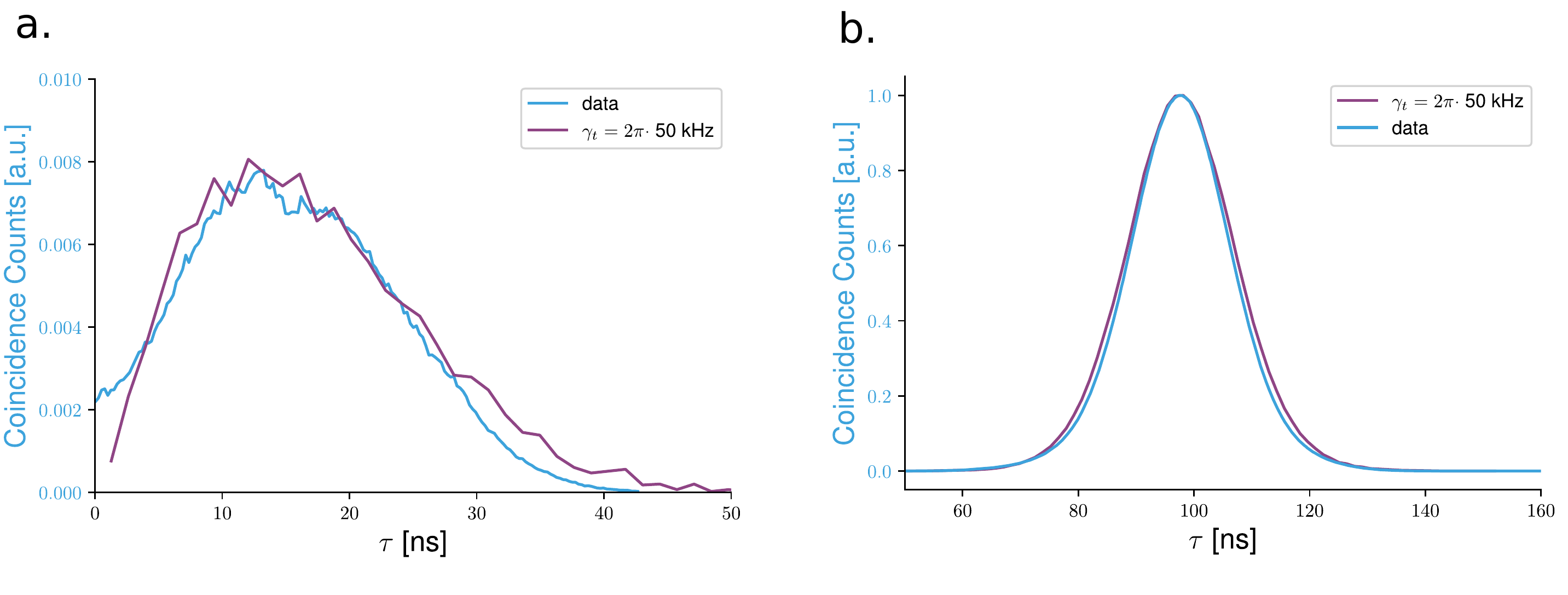}
    \caption{Experimental data and model prediction of the $g^{(2)}$ of Gaussian photon arrivals for the $\tau \approx 0$ region (a) and the first pulse region (b).}
    \label{fig:g2model}
\end{figure}

The second order correlation of photon arrivals, $g^{(2)}$, can be calculated using a slightly modified version of the numerical model. Measuring $g^{(2)}$ requires a re-initialization pulse $\Omega'(t)$ that pumps the population from $\ket{\downarrow, 0}$ to an additional energy level $\ket{\uparrow', 0}$. The population is then restored to the initial state $\ket{\uparrow, 0}$ through a spontaneous decay channel at a rate $\gamma$. Thus, the Hamiltonian becomes
\begin{align}
    \mathbf{H}_{\text{S}}= \Delta \ket{\downarrow^\prime, 0} \bra{\downarrow^\prime, 0} + (\Delta-\Delta_{C})\ket{\downarrow, 1}\bra{ \downarrow, 1} + \Omega(t)\ket{\downarrow^\prime, 0}\bra{ \uparrow, 0} + g\ket{\downarrow, 1}\bra{\downarrow^\prime, 0} +  \Omega'(t)\ket{\uparrow^\prime, 0}\bra{ \downarrow, 0} + \text{h.c.}
\end{align}
and another Lindblad collapse operator $\mathbf{L_4} = \sqrt{\gamma}\ket{\uparrow, 0}\bra{\uparrow', 0}$ is added to the calculation. The master equation can then be solved using the Monte Carlo (MC) method, which keeps track of the times at which a photon is emitted from the cavity ($\mathbf{L_1}$). The MC solver is encoded in the \texttt{solnmc} function of \texttt{QuTip}. Given the pulse sequence of a Gaussian pulse from $\ket{\uparrow, 0}\to\ket{\downarrow', 0}$, a re-initialization pulse from $\ket{\downarrow, 0}\to\ket{\uparrow', 0}$, and another Gaussian pulse from $\ket{\uparrow, 0}\to\ket{\downarrow', 0}$, photon counts occurring after the initial (expected) photon contribute to the $g^{(2)}$-function. Thus, using the MC simulation results, a histogram of photon arrival times can be compiled. In this histogram, we measure photon arrival times relative to the arrival time of the primary photon. These simulations reproduce the experimental $g^{(2)}$-function, as shown in fig. \ref{fig:g2model}.\\

\begin{figure}
    \centering
    \includegraphics[width=0.55\textwidth]{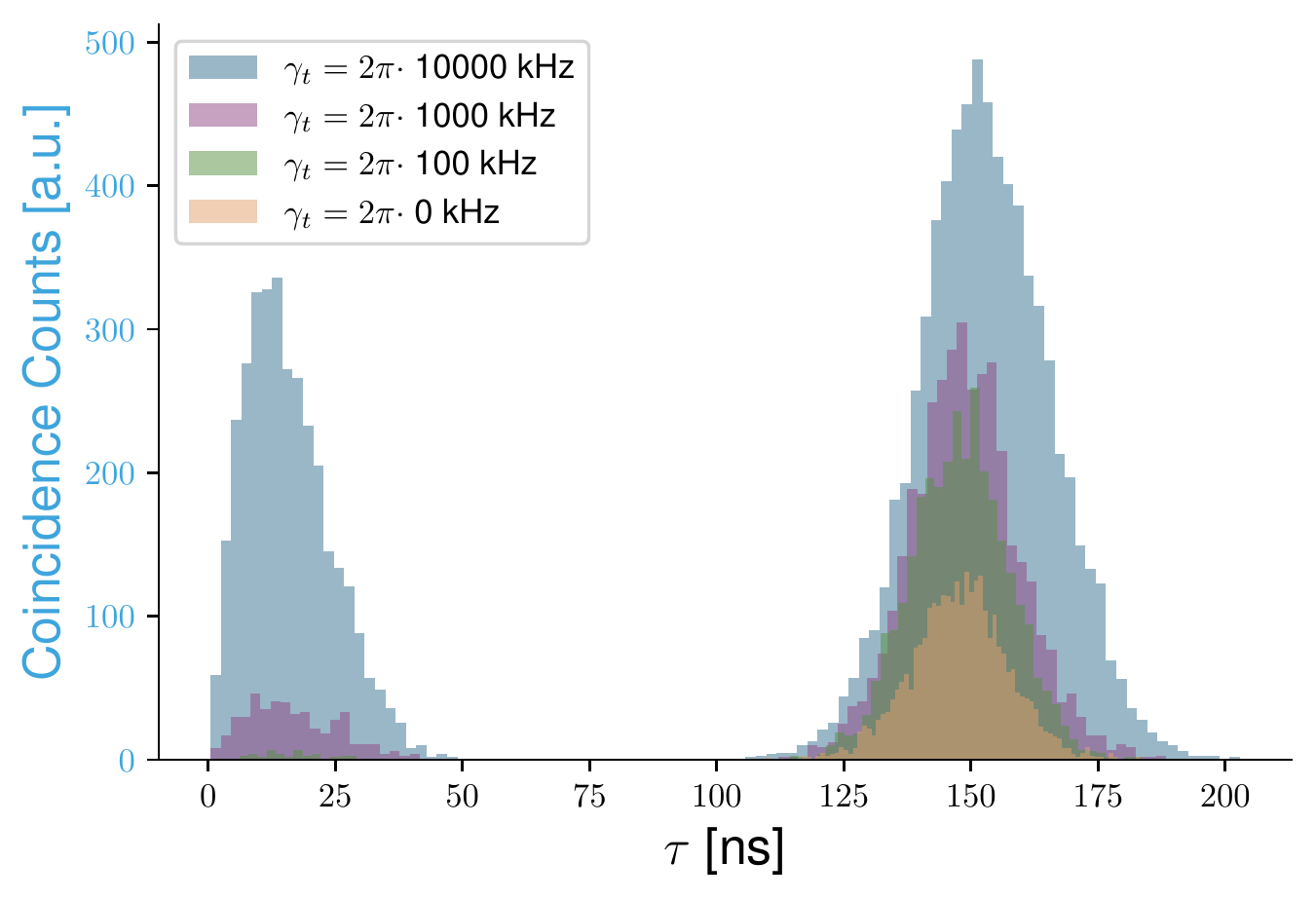}
    \caption{Simulated $g^{(2)}$ functions assuming various qubit decay rates $\gamma_t$ and a control pulse separation of 150 ns. The initial peaks close to $\tau = 0$ are only observable for $\gamma_t \not = 0$ and increase in amplitude for larger $\gamma_t$. This is indicative of emissions of secondary photons due to repopulation. The peaks around 150 ns corresponds to the emission due to the second control pulse and also increases in amplitude for larger $\gamma_t$, which is again attributable to the emission of secondary photons.}
    \label{fig:g2t1}
\end{figure}

The MC simulation can furthermore be used to investigate the influence of the qubit decay rate $\gamma_t$ on the single-photon purity. As shown in \cref{fig:g2t1}, increased values of $\gamma_t$ result in a higher contribution of coincidences for short time delays $\tau \approx 0$. This can be explained by the fact that a high coupling between the two qubit states can result in a re-initialization of the electron in the $\ket{\downarrow}$ state while the control pulse is still applied, which can result in the emission of a photon. Since this is the second photon emitted during the time-window of the first photon, the coincidence counts in this initial time-window around $\tau = 0 $ are increased, and the single-photon purity is decreased. This is supported by our measurements, which show a significantly decreased single-photon purity for the exponentially-shaped photon (main text fig. 3 (a)) and the ten-peaked photon (main text fig. 3 (e)), which both are comparatively long compared to the Gaussian-shaped photon. Furthermore, the coincidence counts close to $\tau = 0$ mirror the temporal shape of the primary photon, which is indicative of a secondary photon emission. As discussed in \ref{section:heating}, we attribute the enhancement of the $\gamma_t$ to instantaneous control laser induced heating.

\section{Experimental Setup}

\subsection{Hardware Overview}

\begin{figure}
    \centering
    \includegraphics[width=0.95\textwidth]{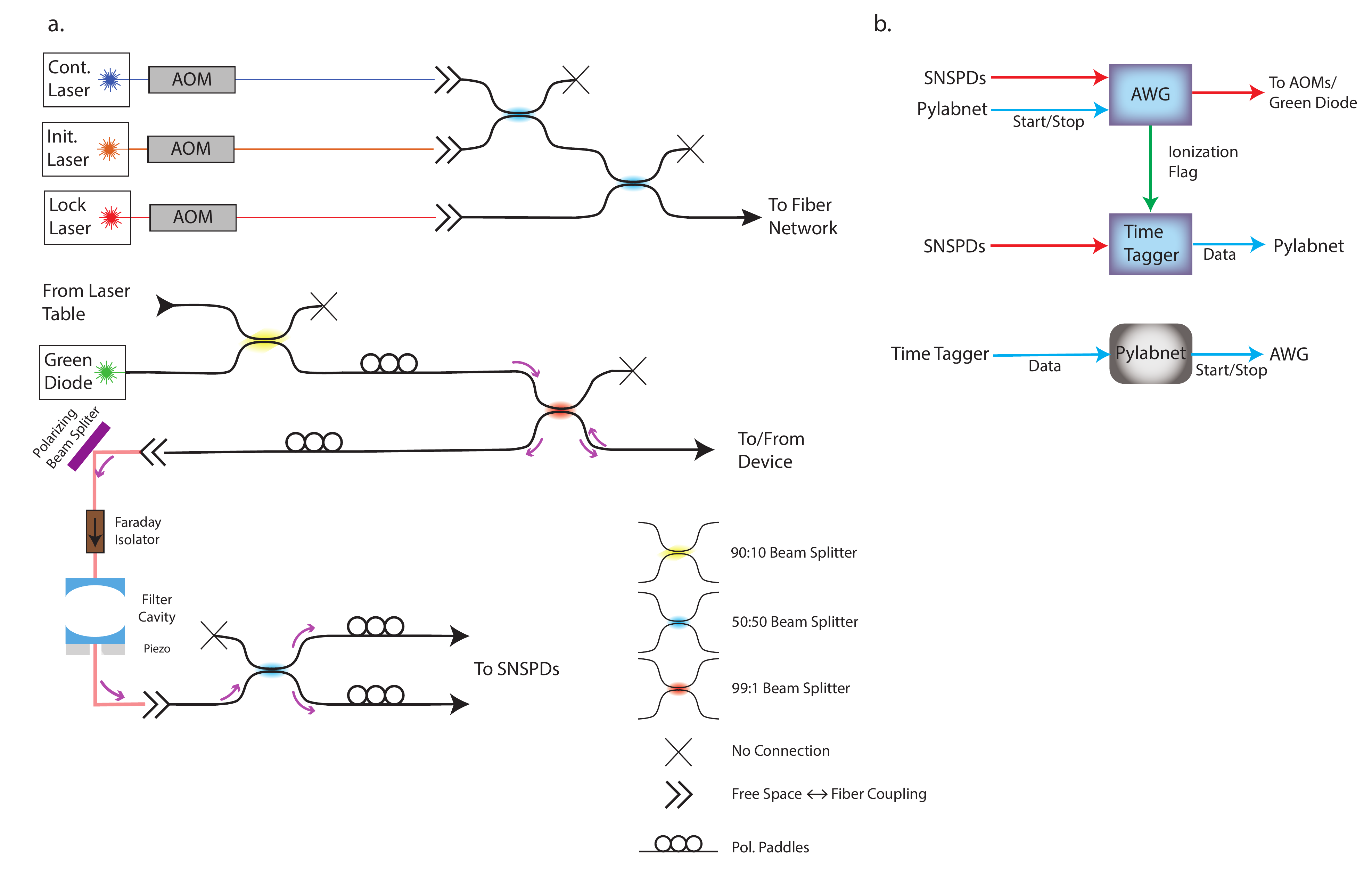}
    \caption{ Diagram of the (a) optical setup and (b) control flow used in this experiment. }
    \label{fig:systemDiagram}
\end{figure}

The full system is detailed in \cref{fig:systemDiagram}. Three primary lasers are used (2 Toptica DLC DL PRO, M Squared SolsTiS-2000-PSX-XF). Two lasers are used to resonantly drive the initialization and photon generation optical transitions, while the third is used to lock the filter cavity. All three were locked to a wavemeter (High Finesse WS7).

The nanophotonic cavity is situated in a dilution refrigerator (DR, BlueFors BF-LD250) with a base temperature of 50 mK. The DR is equipped with a superconducting vector magnet (American Magnets Inc.), a home-built free-space wide-field microscope with piezo positioners (SmarACT SLC-2430 and SLC-1720 series) and fiber feedthroughs. Piezo positioners are used to move the tapered fiber with respect to the diamond sample, and the imaging system is used to facilitate alignment between the tapered fiber and the tapered section of the diamond wave-guide at cryogenic temperatures. This alignment procedure, as well as the subsequently performed tuning of the nanocavity resonance is described in detail in \cite{Evans2018}. The SiV-cavity system is optically interrogated through the fiber network without any free-space
optics \cite{Nguyen2019}.

Pulsing of the lasers is done via acousto-optic modulators (AOMs) controlled by a 2.4 GSa/s arbitrary waveform generator (AWG, Zurich Instruments HDAWG). The digital output lines of the AWG are connected to a custom buffer box to buffer the digital signals, which then modulate the AOMs. 

To detect and process single photon events, superconducting nanowire single-photon detectors (SNSPDs, Photon Spot) are used. The outputs of the SNPSDs are routed to a time tagger (Swabian Instruments Time Tagger Ultra). The time tagger is used not only to count photon detection events but also to perform real time calculations of the cross-correlation of detection events between the two SNSPDs.

\subsection{Software Overview and 24 Hour-Long Run}

In order to run the sequence for generating photons at a high duty cycle and process the data in real-time with a minimum software overhead, a hybrid approach using the custom python suite \texttt{pylabnet} \cite{pylabnet} and logic on the AWG is performed. First, the python program locks the filter cavity to our desired transmission wavelength by turning on (via an AOM) a heavily attenuated lock laser tuned to our single photon frequency and modulating the voltage of the piezo in the cavity until a transmission maximum is achieved as measured by counts on the SNSPDs. The python program then signals to the AWG to run the single-photon generation sequence.

When signaled, the AWG applies the initialization and control pulses to the device via the AOMs. The AWG applies the pulses 250 times for a given sequence. In order to check if the SiV ionized during the applied pulses, a built-in counter gated to the single photon in the AWG is used to count the number of actually detected single photons within that sequence and compare to a minimum ionization threshold. If the number of detected single photons is below the ionization threshold, the SiV is assumed to be ionized, and a green diode laser is applied to the device for $500 \mu s$ to return the SiV to the negative charge state. This enables ionization events to be detected quickly and have minimal impact on the total duty cycle of the source. The number of pulses within a given sequence is optimized to maximize the total duty cycle while still quickly catching ionization events to prevent artificial degradation of the efficiency. This entire sequence is repeated 10000 times before the AWG finishes.

In order to communicate to the time tagger whether the last sequence of data passed the ionization check or not, a digital bit sent from the AWG to the time tagger is used as an ionization flag. The time tagger gates the incoming single photon counts based on the ionization flag, with an appropriate digital delay of approximately a full sequence length added onto the photon click events within the time tagging module so that the AWG will have updated the ionization flag for a given sequence before the events are processed by the time tagger. 

Once the AWG runs the 10000 sequences, control is returned to the python program, which then polls the time tagger for the data, updates the GUI with the new data, and periodically relocks the filter cavity. 

This enables real-time processing of the single photon events and autocorrelation measurements by the time tagging module without significantly impacting the duty cycle.

\subsection{Systemic Loss}

The estimated losses in our system are listed in \cref{tab:tablelosses}. The initialization probability is estimated based on the ratio of the maximum single-photon efficiency achieved to twice the single-photon efficiency when no initialization is performed, where we assume the electronic-hole spin is thermalized to a 50-50 mixed state between each photon run. 

The SiV to waveguide efficiency is calculated based on the Cavity QED parameters (see main text) estimated from fits of the cavity and SiV lineshapes as described previously. Namely, $\eta_{s}= p_{c}*p_{w}$ where $p_{c} = \frac{C}{C+1}$ is the probability of emission of a single photon into the cavity, and $p_{w}=\frac{\kappa_W}{\kappa_{tot}}$ is the probability that a photon in the cavity is coupled out to the waveguide. Note, that $p_{c}$ assumes that the SiV is resonant with the cavity, which is acceptable given the small detuning used in this experiment. (See table \ref{tab:tablesim} for cavity linewidth and detuning.)

The tapered fiber coupling efficiency was measured directly by comparing the reflected power off of the device to the incident power, which is calibrated in the fiber network by substituting the nanophotonic cavity for a fiber coupled retroreflector.

The filtering setup efficiency and fiber network efficiency were measured directly. The uncertainty in the filtering setup efficiency is due to imperfect alignment, drifts in the cavity lock position between cavity relocking sequence, and spectral diffusion of the SiV.

\begin{table*}
\begin{ruledtabular}
\begin{tabular}{ccccc}
 Initialization probability & $0.8-1$ \\ 
 SiV to Waveguide Efficiency & $0.62$ \\
 Tapered Fiber Coupling Efficiency & $0.7$ \\
 Filtering Setup Efficiency & $0.5-0.6$ \\
 Fiber Network Efficiency & $0.92$ \\
 Detector Efficiency & $0.85-0.9$ \\
 \hline
 Overall photon detection efficiency & $0.13 - 0.22$ \\
 \hline

\end{tabular}
\end{ruledtabular}
\caption{ Summary of system losses.}
\label{tab:tablelosses}
\end{table*}

\section{Experimental Results}

\subsection{Single Photon Efficiency Metric}
Two different methods can be used to compute the efficiency of the single photon source. First, the ratio of detection events versus the total number of single-photon generation attempts made (i.e., number of times the control pulse was applied) can be used. Alternatively, as in fig. 4 of the main text, an exponential decay fit to the multi-photon event rates can be used. 

In our experiment, the ratio gives an efficiency of $13.5\%$, whereas the exponential decay fit gives a slightly higher efficiency of $14.9\%$. This discrepancy can be explained by the fact that an ideal exponential decay relationship assumes that each photon detection event is mutually independent and uncorrelated with the other detection events. However, in this system, there is a slow MHz scale diffusion of the SiV qubit splitting over time. Due to the narrow filter cavity, this diffusion results in changes to the detection efficiency. Therefore, there are short term correlations in the detection efficiencies of subsequent photons, making multi-photon detection events more likely than if there was no diffusion. Nevertheless, the efficiency value as determined by the exponential decay fit is the useful metric here because it better represents the multi-photon event rates which is a key figure of merit in photonic quantum information applications.

\subsection{Experimental Duty Cycle}
\label{sec:dutycycle}
Given a control pulse repetition rate of 405 kHz, an average single-photon efficiency of $13.5\%$, and a single photon rate of 31 kHz, an average duty cycle (defined as the portion of time which single photons are actually being generated) of $57\%$ is extracted.

We estimate that $56\%$ of the down-time is due to the ionization of the SiV. This relatively long portion is due to the comparatively long pulse of green required to deionize the SiV ($500 \mu s$). This can be improved in the future, potentially through the use of higher green power or using SiVs less prone to ionization.

Of the rest, we attribute a small portion ($9\%$) to relocking of the Fabry–P\'erot filter cavity. This can be improved through using persistent locking techniques at a detuned wavelength or through optimizing the stability of the cavity and its surrounding environment. The rest ($35\%$) is attributed to the software overhead for the python portion of the experimental software, which is responsible for initially starting the sequence, reading the photon detection data from the time tagging module, and updating the GUI appropriately. This can be further improved through software optimization, particularly of the data transfer between the time tagger and the software through performing local processing of the data prior to transfer to the computer running the GUI.

\subsection{Comparison to Weak Coherent States}
To benchmark the gain from using the single photons generated in this work as opposed to a weak coherent source, we compared the gain in rate when using this source as opposed to a weak coherent source with a 100$\%$ duty cycle and an equivalent repetition rate and two photon-infidelity to our source, where the latter is defined by the ratio between the $n = 2$ and $n = 1$ components of the wavepacket. This gain is given by $D/g^2(0)$, where $D$ is the duty cycle of the source demonstrated here, which gives a 35x enhancement.

To explicitly derive this, we note that for a coherent source, the probability of detection of two photons in the wavepacket is given by $P(n=2 )_{wcs} = \frac{1}{2}P(n=1)_{wcs}^2$ where $P (n=1)_{wcs}$ is the probability of detecting 1 photon in the wavepacket. Thus, we can estimate the infidelity from the higher order component (where we assume the source is weak enough such that higher photon number detection probabilities can be neglected) to be $1-\mathcal{F}_{wcs} = \frac{P(n=2)}{P(n=1)} = \frac{1}{2}P(n=1)_{wcs}$. 

Similarly, for a single photon source with purity $g^2(0)$, the two photon component can be written as $P(n=2)_{sps} \approx \frac{1}{2} g^2(0) P(n=1)_{sps}^2 $ \cite{Senellart2017}, thus giving a corresponding infidelity of $1-\mathcal{F}_{sps} = \frac{P(n=2)}{P(n=1)} = \frac{1}{2}g^2(0)P(n=1)_{sps}$. We again note that we assume that higher photon component probabilities are negligible. 

We now observe that setting $P(n=1)_{wcs}$ so that the infidelity of the weak coherent source is equal to that of the single photon source ($\mathcal{F}_{wcs} = \mathcal{F}_{sps}$), requires that $P(n=1)_{wcs} = g^2(0)P(n=1)_{sps}$.

From this, if we assume a weak coherent source operated at the same repetition rate as the single photon source, a single photon rate enhancement of our single photon source versus a comparable weak coherent source could be estimated as $\frac{P(n=1)_{sps}}{P(n=1)_{wcs}} = 1/g^2(0)$. However, in order to provide a fair comparison, we note that our single photon source has a non-unity duty cycle $D$ due to a variety of experimental imperfections (see \ref{sec:dutycycle}), whereas as typical coherent sources can be operated continuously. Thus, a better estimation of the rate enhancement provided by the single photon source is given by $D\frac{P(n=1)_{sps}}{P(n=1)_{wcs}} = D/g^2(0)$. From this expression, we estimate the 35x gain reported in the main text.

\section{Limitations}
\subsection{Si$^{29}$ Readout Limited Lifetime}
In the main text, a decay of the nuclear spin population after photon generation is observed through the photon autocorrelation measurements. The observed nuclear spin lifetime is significantly shorter than the expected. This suggests a shortening of the nuclear spin lifetime due to the photon generation process.

\begin{figure}
    \centering
    \includegraphics[width=0.5\textwidth]{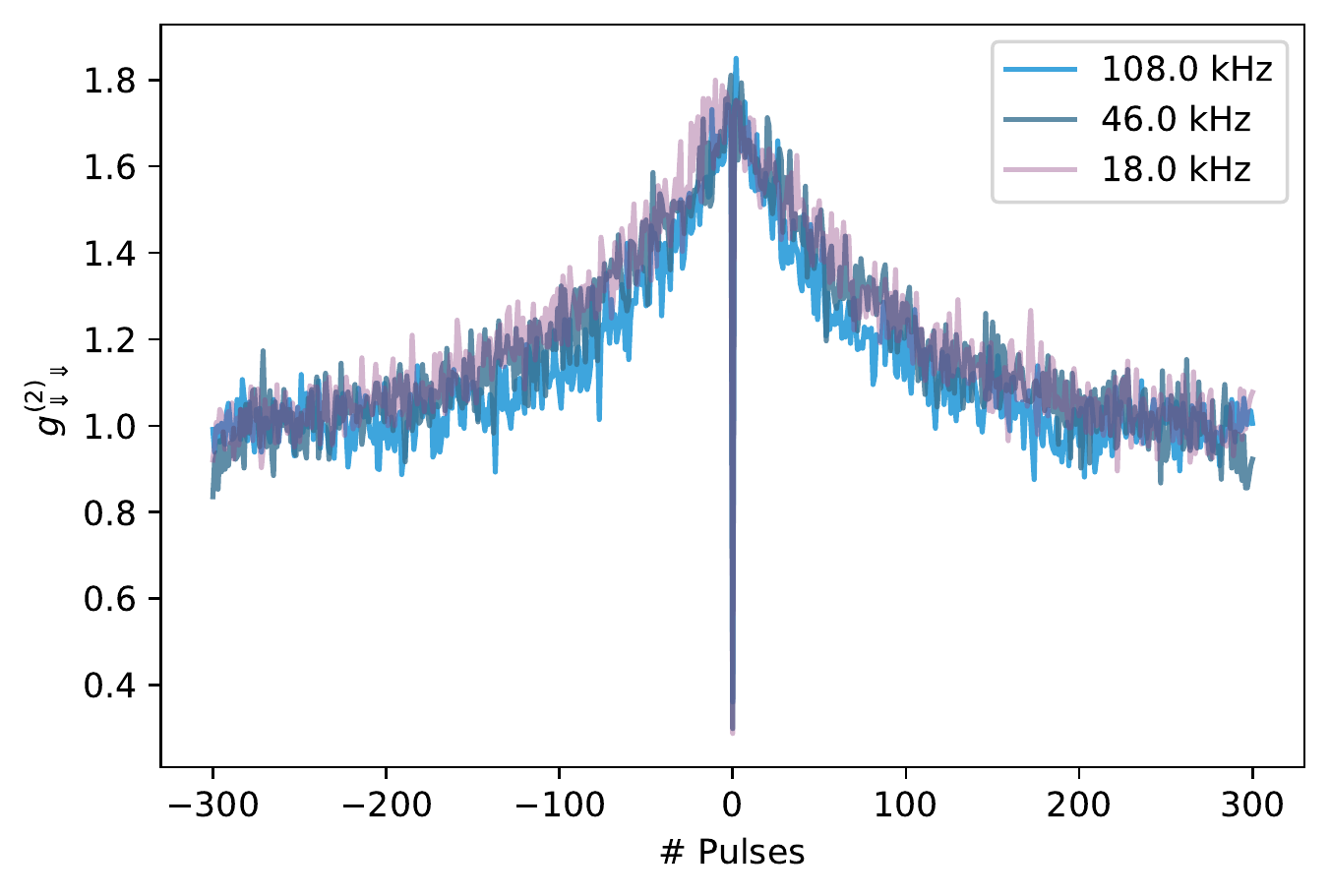}
    \caption{ ${g_{\Downarrow\Downarrow}}^{(2)}$ auto-correlation measurements of the photons emitted at frequency $\nu_{\Downarrow}$ at varying repetition rates, showing an approximately constant bunching time scale.}
    \label{fig:nucRepRate}
\end{figure}

To further verify this, the photon bunching curve is measured at varying photon generation repetition rates (\cref{fig:nucRepRate}). We see that despite the varying time between generated photons involved, the bunching decay constant is more or less constant as a function of the number of photons. This indicates that the photon generation process itself is inducing the nuclear spin flip. 

The exact mechanism for this shortening of the nuclear spin lifetime via the generation of photons is still under investigation. Despite this, we note that the number of scattered photons is significantly higher than the largest measured multi-photon stream, and thus in the near term, the limit to generating large multi-photon entangled states will be efficiency as opposed to the lifetime of the nucleus. 

\subsection{Heating Limited Electron T$_1$}
\label{section:heating}
\begin{figure}
    \centering
    \includegraphics[width=0.85\textwidth]{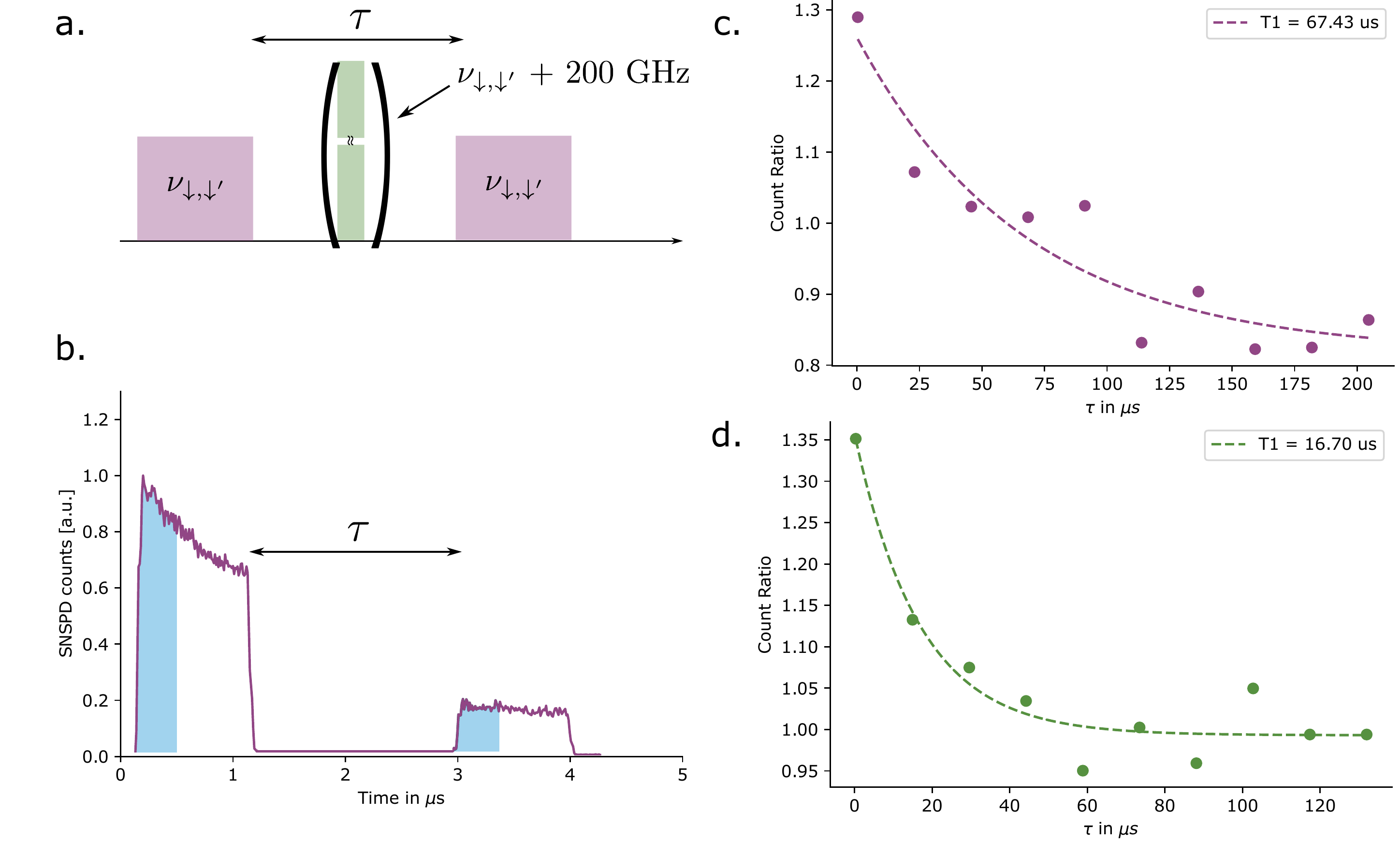}
    \caption{(a) Pulse sequence used for T$_1$ measurements. A pulse along the optical spin-conserving transition optically pumps the electron spin from the $\ket{\downarrow}$ to the $\ket{\uparrow}$ level. After a dark period $\tau$, the same pumping pulse is applied. An optional strong off-resonant pulse is applied to emulate the heating induced by the strong control pulse $\Omega_{\text{control}}$. (b) Exemplary time trace resulting from the pulse sequence in (a). To probe relaxation from $\ket{\uparrow}$ to $\ket{\downarrow}$, the ratio of the two shaded regions is plotted for different dark times $\tau$ both assuming no heating pulse (c) and with applied heating pulse (d). The spin-lifetime T$_1$ is extracted by fitting the obtained data to an exponential proportional to $\exp(- t / \text{T}_1)$.}
    \label{fig:t1meas}
\end{figure}

The photon generation scheme relies on applying a strong control pulse $\Omega_{\text{cont}}$ along the weak spin-flipping transition. Our nanophotonic cavity is embedded in a free-standing waveguide. It was shown previously that this free-standing structure is prone to microwave-driving induced heating \cite{Nguyen2019a}. In order to analyze the impact of the strong control pulse $\Omega_{\text{control}}$ on the lifetime of the excited state $\ket{\uparrow}$, we first carry out a optical-pumping based lifetime measurement \cite{Jahnke2015}. This is done by applying a pulse along the $\ket{\downarrow'} \rightarrow \ket{\downarrow'}$ transition, which optically pumps the electron into the $\ket{\downarrow}$ state (\cref{fig:t1meas} (a)). After a dark time $\tau$, this optical pumping pulse is repeated. The timetrace of the SNSPD counts during this sequence (\cref{fig:t1meas} (b)) show a fluorescence peak, which decays over the course of the pulse. This decay is indicative of the optical pumping from $\ket{\downarrow}$ to $\ket{\uparrow}$. If the electron spin has relaxed back to $\ket{\uparrow}$ during the dark period, the second application of the optical pumping pulse will show a similar decaying fluorescence feature. Thus, by plotting the ratio of the counts of the first and second optical pumping pulse, the lifetime of the excited state $\ket{\uparrow}$ can be extracted (\cref{fig:t1meas} (c)). \\

To analyze the impact of a strong control pulse $\Omega_{\text{control}}$ on the electron lifetime, we re-run the T$_1$ measurement while applying a strong pulse during the dark time. This pulse is identical to the $\Omega_{\text{control}}$ pulse used to generate single photons with the exception of the frequency, which is detuned from any optical SiV transition. This is done to ensure that the strong pulse does not resonantly interact with the SiV. We can see that upon application of this strong pulse, the measured T$_1$ decreases more than three-fold. This decrease in T$_1$ persists even if the repetition rate is lowered by an order of magnitude. Since we can exclude resonant interactions, we attribute this decrease in T$_1$ to instantaneous heating induced by the control laser pulse. During single-photon generation, this decrease in T$_1$ can be overcome by ensuring the single photon is generated on timescales faster than T$_1$ (see discussion in \ref{section:g2sim}). For application where radio-frequency control of the $^{29}$Si nucleus is needed, this heating induced deterioration of T$_1$ will need to be overcome in order to allow for nuclear operations which extend over multiple $\mu s$.

One strategy to mitigate the SiV's susceptibility to heating induced $T_1$ reduction is to work with strained emitters, which can show an increased ground-state splitting $\Delta_{\text{gs}}$. Large values of $\Delta_{\text{gs}}$  result in a decreased probability of phonon-induced depopulation and can increase $T_1$ times by an order of magnitude \cite{srujanstrain}. The necessary strain can either be obtained by pre-selecting suitable emitters, or by integrating nanophotonic cavities into previously demonstrated strain tuning devices \cite{Machielse2019}. 
Alternatively, the heat load introduced by radio-frequency driving of the nucleus can be reduced by re-engineering the co-planar waveguide (CPW) used to drive the SiV nucleus. Previously, gold CPW's have been used to deliver radio-frequency pulses to the SiV \cite{Nguyen2019}. Changing the CPW material to a suitable superconducting material could result in lower losses at radio frequencies, which would reduce the heat load on the emitter.

\section{High Temperature Operation}
The ability to generate single photons with the SiV at elevated temperatures ($\sim1$K) would significantly relax the cryogenic experimental overhead required for operating the source. Normally SiV quantum memories cannot be operated at these temperatures due to their susceptibility to phonon processes. Interestingly, the photon generation protocol demonstrated here is fairly immune to these requirements due to the ability to generate short photons, and thus, theoretically, should be operable at 1K.

There are three main sources of degradation of the SiV's spin properties at 1K. First, the T$_2$* of the spin levels significantly degrades. We estimate it to be around $\sim400$ ns based on extrapolation from data from \cite{Pingault2017}. With generated photons with a width of $\sim20$ ns, and assuming a Markovian-limited T$_2$*, a small increase of the linewidth $2.5\%$ is expected based on the convolution of the spectral intensity of the Gaussian photon and the dephasing. The second potential pathway for degradation is the drop in the spin lifetime, extrapolated to be $\sim1.2\mu$s. Given our photon width of 20 ns, we would expect approximately a population decrease of $\sim 2\%$, which would degrade the efficiency by that amount and also potentially limit the $g^2(0)$ similarly. Finally, we expect an increased population trapped in the upper branch of the ground state manifold, which would decrease the efficiency of the source. Given a ground state splitting of 46GHz, we would expect a population of $\sim 11\%$ in the steady state at 1K, which is the largest impact of operating at higher temperatures but is not prohibitive and could be improved via optically pumping the upper branch during initialization.

Thus even given the degraded spin properties, we would still be able to generate bandwidth-matched $\sim 20$ ns photons at 1K, with only a small degradation in efficiency, highlighting the versatility of this protocol.

\bibliographystyle{apsrev4-2} 
\bibliography{refs}